\documentclass{sig-alternate-10pt}

\usepackage{url}

\usepackage{algorithm}
\usepackage{algpseudocode}
\usepackage{topcapt}
\usepackage{booktabs}
\usepackage{color}

\newcommand{\mpara}[1]{\medskip\noindent{\bf{#1}}}

\usepackage{multicol}

\usepackage{mathptmx}

\usepackage{amsmath}

\usepackage{amssymb}

\usepackage{graphicx}

\newcommand{\calP}{\ensuremath{{\mathcal{P}}}}

\newcommand{\RSD}{\mbox{\em RSD}}
\newcommand{\rsd}{\mbox{\em rsd}}
\newcommand{\rsim}{\mbox{\em rsim}}
\newcommand{\dg}{\mbox{\em dg}}
\newcommand{\ag}{\mbox{\em ag}}
\newcommand{\nexthop}{\mbox{\em nexthop}}
\newcommand{\routestate}{\mbox{\em routestate}}
\newcommand{\RSclustering}{{\sc RS-Clustering}}

\newcommand{\pivot}{{\tt Pivot}}

\newcommand{\PartialSpearmanFootrule}{{\sc Partial Spearman Footrule Distance}}
\newcommand{\Geometric}{{\sc Geometric Distance}}
\newcommand{\psdist}{\mbox{\em ps-dist}}
\newcommand{\gdist}{\mbox{\em g-dist}}

\newcommand{\dsjuly}{{\tt Day-1}}
\newcommand{\dsjan}{{\tt Day-2}}

\newif\ifnotes
\notesfalse
\ifnotes
\newcommand{\authnote}[2]{{ $\ll$\textsf{\footnotesize #1: #2}$\gg$}}
\else
\newcommand{\authnote}[2]{}
\fi

\begin{document}
\thispagestyle{empty}


\title{Routing-Aware Partitioning of the Internet Address Space for Server Ranking in CDNs \\ \small{(A more recent version of this manuscript is published in Elsevier Computer Communications, vol. 106, July 2017)}}

\author{
\centerline{{Gonca G\"ursun}}\\
\centerline{\sc{Department of Computer Science, Ozyegin University}} 
}
\maketitle

\begin{abstract}
The goal of Content Delivery Networks (CDNs) is to serve content to end-users with high performance. 
In order to do that, a CDN measures the latency on the paths from its servers to users and then 
selects a best available server for each user. For large CDNs, monitoring paths from 
thousands of servers to millions of users is a challenging task due to its size. 
In this paper, we address this problem and propose a framework to scale the task of 
path monitoring. Simply stated, the goal of our framework is clustering 
IP addresses (clients) such that in each cluster the choice of best available server is same (or similar).
Then, finding a best available server for one client in a given cluster will be sufficient to assign
that server to the rest of the clients in the cluster.

To achieve this goal, first we introduce two distance metrics to compute how similar 
the server choices of any given two clients. Second, we use a clustering method that 
is based on interdomain routing information. We evaluate the goodness of our clusters 
by using the metrics we introduce. We show that there is a strong correlation between 
the similarity in how two destination clients are routed to in the Internet and the similarity in 
their server selections. Finally, we show how to choose representative 
clients from each cluster so that it is sufficient to learn the latencies from the CDN 
servers to the representative and find a best available server accordingly for the rest of the 
clients in the same cluster. 


\end{abstract}

\section{Introduction}    \label{sec:intro}   A Content Delivery Network (CDN) is a collection of servers that deliver content 
to end-users on behalf of content owners. Today, significant amount of Internet traffic 
is served by CDNs. For example, one of the largest CDNs, Akamai, currently delivers 
15-30\% of all web traffic from a large distributed platform. This platform consists of 
over 160,000 servers in more than 100 countries and 1200 ISPs around the 
globe \cite{vis-akamai}. 

The main job of a CDN is replicating content across its geographically distributed 
server regions and redirect end-users to a \emph{best available} server region 
at a given time. The goal is serving end-users with high performance 
\cite{Dilley02}, that is each user is redirected to a region to 
which it has low latency. One can expect that mapping an end-user 
to its geographically closest server region is sufficient. 
However, there are many cases where geographical closeness does
not infer low latency \cite{Krishnan09}. Instead, the conditions in 
the network determine which server regions are best performing at a given time.
The challenge is, since the network is dynamic, the conditions on its paths
are to change frequently. Therefore, the CDN needs a monitoring solution
that can keep pace with the variability of the Internet paths. 

In large CDNs, such as Akamai, monitoring paths between all the server
regions and end-users is challenging due to the scale of the task. 
At a given day Akamai sees more than 788 million unique IPv4 
addresses \cite{state-akamai}. It is not feasible to measure 
latencies from hundreds of thousands of servers to all these
IPs. In the practice of traditional DNS-based mapping, end-users are
represented by their local DNS resolvers and the path measurements are
taken between the servers and the local DNS resolvers \cite{Nygren10}. 
Although, this reduces the size of the task up to some extent, there is 
still need for clustering local DNS resolvers since there are millions of 
them in the Internet. Recently, with the increase in usage of public DNS 
resolvers \cite{google-dns, open-dns} and the adoption of EDNS \cite{Streibelt13}, 
CDNs move towards end-user mapping \cite{Chen15}. In the case of end-user
mapping, the users are not represented by their local DNS any more. That is, the need for partitioning 
the Internet address space to scale the path monitoring task is more crucial than ever. 
In this paper, we present a method that reduces the size of the path monitoring and 
server ranking tasks both in the case of DNS-based and end-user based mapping.


Our study has the following three stages.

\mpara{1. Clustering clients.} We seek to find a partitioning of 
the IP address space such that the clients 
in a given partition\footnote{Throughout this paper, we use the terms \emph{clustering}
and \emph{partitioning} interchangebly.} 
orders the server regions from least latency to most 
in similar fashion. The reason why we are not only interested in finding 
the least latency region but also in ordering the regions is as follows. 

In addition to high performance, \emph{best available} server region 
is subject to some other constraints such as load balancing at the CDN,
availability of the requested content at the server, allowance rules
(enforced by ISPs and content publishers) on serving specific contents 
to users from specific regions etc. Therefore, for a given client, a best 
available server region is the one with the lowest latency that also satisfies
the constraints. For that reason, the clients \emph{rank} 
the server regions from least latency to most and then these rankings are 
used as input to the server mapping algorithm.

The clustering scheme we propose for our problem is called 
\RSclustering\ and it is introduced in \cite{Gursun12}. \RSclustering\
is a method that groups BGP prefixes based on how similar
the ASes in the Internet route to these prefixes. The key idea behind 
using this clustering scheme for our problem stems from the fact that 
routing is one main factor that impacts path latencies.
Therefore, our hypothesis is that if traffic from the server regions 
to two client prefixes follow the same paths, then these clients 
rank the server regions similarly. We show that this intuition holds. 
Routing-aware clustering successfully partitions the address space and outperforms
other clustering methods, such as the ones based on AS or geography. 


\mpara{2. Evaluating the Goodness of Clusters.} Once the clusters
are obtained, the next step is evaluating their goodness. 
In a good cluster, we expect that the server rankings of clients to be similar 
to each other. Such similarity can be defined in various ways. For instance,
one can expect that in a good cluster, all clients have the exact 
same server region as their first (rank-1) choice. Or alternatively, one can 
expect that each server region is ranked in close positions by all clients in 
the cluster. 

To capture these expectations we propose two metrics, 
called \Geometric\ (\gdist) and \PartialSpearmanFootrule\ (\psdist) 
We use \gdist\ and \psdist\ to measure the similarity between two
server rankings. Using these metrics we show how to evaluate any
given clustering scheme. 

\mpara{3. Finding representative clients for each cluster.} Finally, 
we seek to find a client from each partition whose server ranking is a 
good representative of all the other clients in its partition. We scale the task of 
path monitoring by taking measurements only to the representative of
each cluster. 
We first find a \emph{consensus ranking} per cluster by aggregating 
the rankings of all clients in the cluster. Then, we show that assigning one 
client at random from the center of the cluster is almost as good as the
consensus ranking. 


\mpara{Roadmap.} The rest of the paper is organized as follows. 
In Section~\ref{sec:background} we describe the server mapping 
system of Akamai. In Section~\ref{sec:metric}, we introduce the 
metrics we use to evaluate the goodness of clusters and follow by  
describing our dataset in Section~\ref{sec:dataset}. In Section~\ref{sec:pfx-eval}, we set basis for the routing-aware 
clustering by investigating whether IP addresses can be pre-clustered to their 
BGP prefixes. In Section~\ref{sec:clust-eval} we show how to group BGP prefixes further
based on inter domain routing choices in the Internet. 
In Section~\ref{sec:application} we show how to find representative nodes per cluster
to scale the path monitoring task.
We present related works in Section~\ref{sec:rel-work},
discuss some issues related to our work in Section~\ref{sec:discussion}, and finally
conclude in Section~\ref{sec:conc}.

\section{Background}   \label{sec:background}  In this section, we first provide a high-level description of the server mapping system in Akamai's CDN. 
Next, we present the challenges in the system and the goals of our work. 

\subsection{Ranking Server Regions}
\label{subsec:scoring}

\begin{figure}[tbp]
\centerline{
\includegraphics[width=0.5\textwidth]{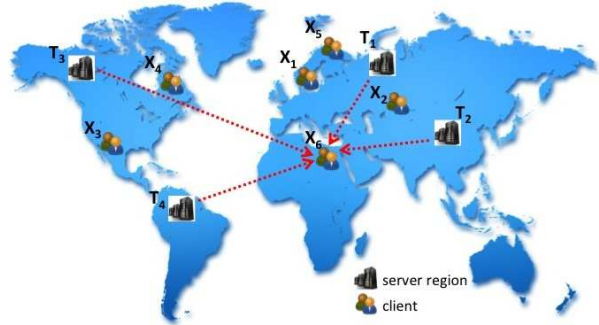}}
\caption{Simplified view of Akamai CDN architecture with four Akamai server regions, $ \{T_1, T_2, T_3, T_4 \}$, 
and six clients $\{X_1, X_2, X_3, X_4, X_5, X_6\}$. Dotted lines show the paths that are monitored for client $X_6$.}
\label{fig:worldmap}
\end{figure}

\begin{figure}[tbp]
\centerline{
\includegraphics[width=0.24\textwidth]{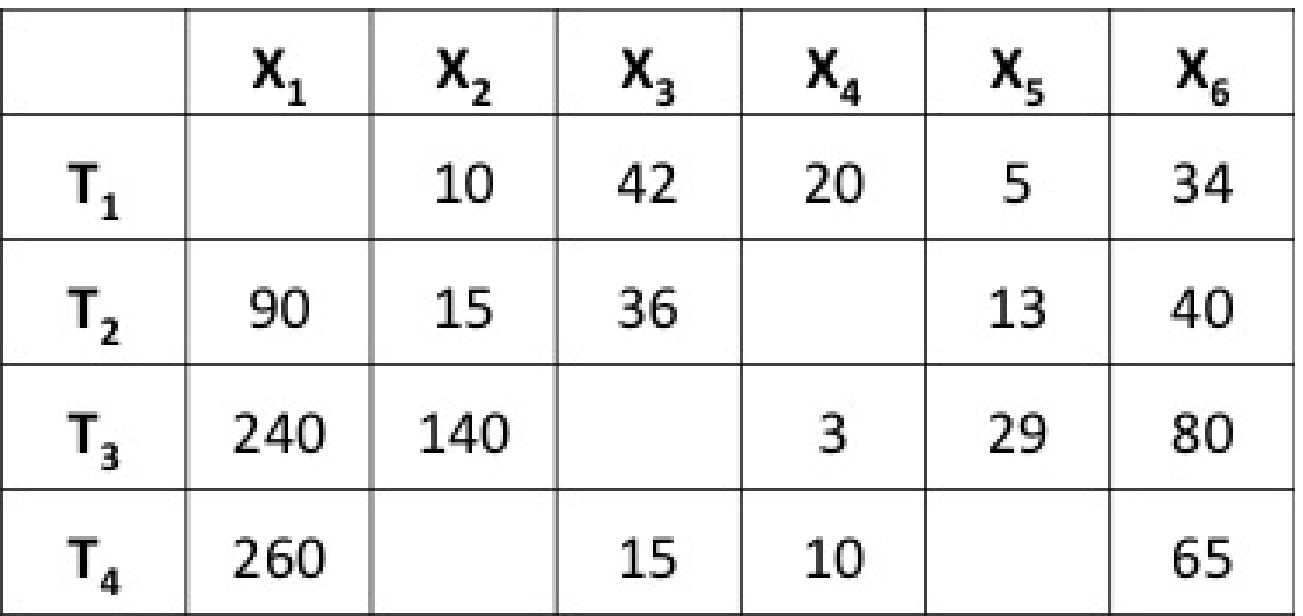}
\includegraphics[width=0.24\textwidth]{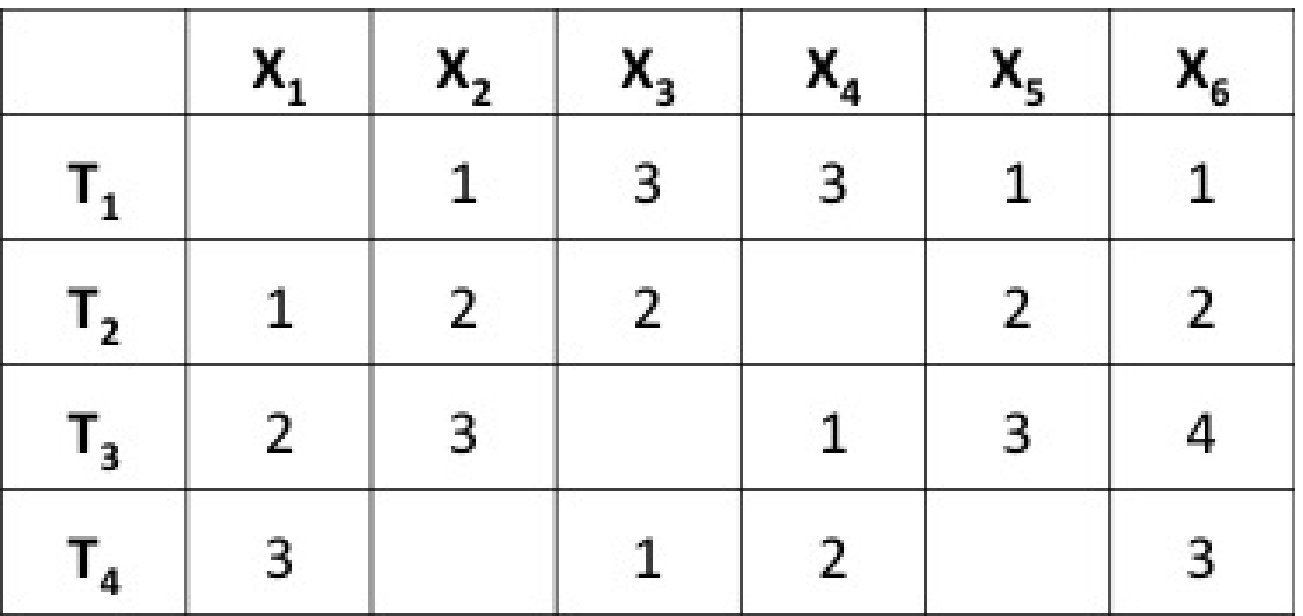}}
  \centerline{\hfill\hfill(a)\hfill\hfill\hfill\hfill\hfill(b)\hfill\hfill\hfill}
\caption{Latencies between the server regions and the clients (a) and the corresponding rank vectors (b) for the example in Figure~\ref{fig:worldmap}.} 
\vspace{-1.2em}
\label{tab:example}
\end{figure}

The core component of Akamai's CDN is the mapping system. One main job of the mapping system
is finding a \emph{best available} server region for each client. In order to do that, first, the mapping system 
monitors the network conditions in the Internet to learn the latencies on the paths from the server regions 
to the clients. Next, using these measurements, it generates a list per client by ranking the server 
regions from best performing to worst. Finally, these candidate lists are used as input to the 
server region assignment algorithm that matches one server region with each client. 

Note that in practice server region assignment algorithm has various other inputs and constraints in addition to latency. 
For instance, delivery cost of traffic (that varies based on ISPs and the content), availability of the content in the servers, 
capacity of the servers, the allowance rules enforced by ISPs, the type of the content application 
are some of the constraints that effect the choice of a best available
server for a given client. That is, a best available server for a client is not necessarily the least latency one.
In fact, this is exactly the motivation behind ranking the servers instead of just finding the best performing server, i.e.\
the best performing one might not be the best available match in terms of the additional constraints. 
In practice, these constraints are applied after the ranking is performed.
Their use and how the server matching is performed based on these constraints are not in the scope of this paper.
Instead, we study the methods that lists the server rank lists based on path performance in a scalable fashion.

Figure~\ref{fig:worldmap} shows a simple example with six clients and four Akamai server regions. 
Both Akamai server regions and the clients are spread across the globe. For each client, the paths from 
server regions to the client are monitored as shown for client $X_6$ in the example.
Let $ \{34, 40 , 80, 65\} $ be the latencies in milliseconds from the set of server regions $ \{T_1, T_2, T_3, T_4 \}$
to $X_6$, respectively. Then the ranking list of $X_6$ is $x_6 = \{1, 2, 4, 3\}$.

At the high-level, there are two main challenges in this context. First, due to the dynamic nature of the Internet, 
the mapping system works in real-time. Therefore rank lists 
need to  be regenerated a few times in a minute. This requires constantly measuring path performance. Second, the scale of 
the problem is too large. There are millions of end-users and taking measurements from thousands of server regions to 
each of them is not feasible. Note that in the case of DNS-based mapping, end-users are represented by their DNS 
resolvers, i.e. measurements are taken between the server regions and the DNS resolvers to estimate the 
path performance to end-users behind each DNS resolver. Even in that case, the scale of the problem is too 
large since there are millions of DNS resolvers \footnote{Note that methods described in this paper
are applicable to not only end-users but also DNS resolvers. Therefore, in the rest of the paper we will use the term \emph{client} to refer to either \emph{DNS resolver} (in the case of DNS-based mapping) 
or \emph{end-user} (in the case of end-user mapping)}.

To overcome these challenges and make the monitoring scalable, the mapping system aims to cluster clients into 
groups and select one \emph{representative} client per group such that
measurements are taken for only the representative client. Then, all clients in the group are assigned to a server based on the ranking of the representative client.

\subsection{Goals and Challenges}
\label{subsec:goals}

Our first goal in this work is finding a partitioning of the Internet address space so that, 
in each partition, the server region rankings of the clients are similar to each other. We
expect a good partitioning to be stable, i.e.\ the clients do not migrate from one partition to the other
one frequently over time. There are various ways of partitioning the address space, e.g.\ by mapping 
clients to their geography or autonomous systems. Although these mappings might work up to 
some extent, our intent is a partitioning that is driven by the network dynamics in order to capture the 
changes on the paths. For that reason, we propose using a clustering method that is based on 
the routing state in the Internet.

Our second goal is developing metrics to evaluate the goodness of a given partition. That is, given
a pair of clients, our intent is quantifying their similarity in terms of ranking server regions.
There are well-known metrics such as Kendall tau and Spearman footrule that computes the correlation
between two rank vectors \cite{Diaconis88}. The assumption behind these correlation metrics is that
the rank vectors are fully known, i.e.\ each region is assigned a rank position by each and every client. 
However, in practice, even for a single client, taking measurements 
from all the server regions to the client is not scalable. Therefore, the mapping system samples 
from the set of server regions for each client 
and take measurements only from this sampled subset. Such sampling is necessary in order not to 
overwhelm the clients with large number of requests. 
This results in learning the performance of the paths from only a subset of server regions per client. 
The subset vary for each client. For instance, 
in Figure~\ref{fig:worldmap}, assume that measurements from only 3 server regions are available for some clients. 
Then, the known latencies and the corresponding rankings are as shown in Figure~\ref{tab:example} (a) and (b),
respectively. The empty entries in the tables are due to the unknown latencies. 
The missing values in the rank vectors make the well-known rank correlation metrics unsuitable.
To overcome this challenge, in the next section, we propose metrics that measure similarity between 
server rankings in the case of unknown measurements.

Note that the actual server assignment algorithms is not in the scope of this paper. 
Instead, we study the server ranking problem which is an input to the server assignment algorithm. 
To that end, our goals are 1) finding a method that clusters IP addresses based on similarity in path performance they receive
from server regions, 2) developing the metrics that evaluate the success of clustering, and 3) developing 
methods to select representative clients to scale the server ranking problem.

\section{Evaluation Metrics}    \label{sec:metric}   Before we introduce our clustering method, we first propose two metrics 
to measure the similarity between two clients based on
how they rank server regions. In the following sections, we use these 
metrics to evaluate how good (compact) clusters are.
 
A successful partitioning generates clusters where 
the server rankings of clients within a given cluster are similar to each other. 
We propose two similarity definitions: 1) we expect that in a good cluster, 
server regions are ranked in close positions. In other words, two rankings that
are close to each other in $l_1$ norm should be grouped together. 
2) top few rank positions are more important than the others since they have higher probability to be 
matched with the client. Therefore we expect that in a good cluster, all users have same server regions 
for their top few rank positions. That is, each rank position has a weight that is proportional to its 
order in the ranking. Note that this is a much more strict 
constraint than the previous definition and more sensitive to the missing measurements. 

To capture each of these similarity definitions we propose two metrics, 
called \PartialSpearmanFootrule\ (\psdist) and \Geometric\ (\gdist).





Let $x_1$ be a real-valued rank vector of length $m$ for client $X_1$,
where $m$ is the total number of server regions. 
We define a ranking (or ordering) of its elements $\sigma_1$ such that
$\sigma_1(i)$ is the rank position of element $i$ in the sorted 
$x_1$. That is $\sigma_1$ is a permutation of numbers from 1 to $m$.
We say that in $x_1$, $i$ is preferred over $j$ if $\sigma_1(i) < \sigma_1(j)$.

Given two vectors $x_1$ and $x_2$ of same length, below we define two distance metrics 
between their ranking vectors $\sigma_1$ and $\sigma_2$, respectively. 

\subsection{Partial Spearman's Footrule Distance}
\label{sec:psdist}

One well-known distance metric for rankings is Spearman's footrule. 
It is the $l_1$ distance between two vectors $\sigma_1$ and $\sigma_2$ s.t.  
$\rho = \sum_{i=1}^m | \sigma_1(i) - \sigma_2(i)| $ \cite{Diaconis88}. 
By definition, Spearman's foot rule distance is maximum when the ordering in $\sigma_1$
is the reverse of the ordering in $\sigma_2$. 

Note that Spearman's footrule requires the complete information on the rankings, i.e. 
for each entry $i$ of $x_1$ the ranking of $i$ must be known (likewise for $x_2$). In cases where 
$x_1$ and $x_2$ are only partially known, Spearman's foot rule can be modified as 
follows \cite{Fagin03}.  

Let $\tau_1$ be the ranking of elements in $x_1$ such that $\tau_1(i)$ is the rank 
position of element $i$ only if $x_1(i)$ is known. Let $k$ be the 
number of known elements in $x_1$. Then for any unknown element $j$, one can set its 
rank to $l$, where $l > k$. Likewise $\tau_2$ is defined for $x_2$.

The intuition is that the servers whose latency are unknown 
are still considered (equally) but not preferred over the servers whose latency are known.
This modified version of Spearman Footrule Distance have some nice properties
(e.g.\ being a metric) as discussed in \cite{Fagin03}. 

Then, the Partial Spearman's footrule distance between $x_1$ and $x_2$ is  
$ PS-dist(x_1, x_2 ) = \sum_{i=1}^m | \tau_1(i) - \tau_2(i)| $.

In this work, we set all the unknown ranks to $l = k + 1$. 
Then, we normalize $PS-dist$ by $k\times(k+1)$ so that it is always between 0 and 1. 
We call the normalized distance as \psdist.

For the example in Figure~\ref{tab:example}, $k = 3$.  Therefore any unknown rank value 
is assigned 4. Then, \psdist$(x_2, x_4) = \frac{8}{12}$ and
\psdist$(x_2, x_3) = \frac{6}{12}$. 


Note that \psdist\ computes the distance between two rankings without assigning weights to the rank positions. 
However, in some applications the higher rank positions matter more than the lower ones.  
Below we define another distance metric that assigns weights to the rank positions such that
the distance between $x_1$ and $x_2$ is smaller when the elements in the higher
rank positions are the same.

\subsection{Geometric Distance}
\label{sec:gdist}

We define geometric distance (\gdist) between $x_1$ and $x_2$ as follows:

\begin{equation}
	\gdist(x_1, x_2) = 1 - \sum_{i=1}^m {I(i) \frac{1}{2^i}}
\end{equation}

where $I$ is an indicator function s.t. $I(i) = 1$ if $x_1$ and $x_2$ both prefer the same element 
for the $i^{th}$ position, otherwise it is 0.

Note that $\gdist(x_1, x_2) \to 1$ as $m \to \infty$ and it drops proportionally to the importance of the rank position. 
For instance if the highest ranked element (i.e.\ rank-$1$) of $x_1$ and $x_2$ are the same then their distance 
is guaranteed to be less than or equal 0.5. Likewise, if both their rank-$1$ and rank-$2$ elements are the same 
then their distance is guaranteed to be less than or equal to 0.25. For the example in Figure~\ref{tab:example}, 
\gdist$(x_2, x_4) = 1$, \\
and \gdist$(x_1, x_2) = \frac{1}{2}$. 

One very important point to note is that we use neither \psdist\ nor \gdist\ to 
cluster the clients. We only use them to evaluate the goodness of an already 
formed cluster. Although, our purpose is grouping clients whose server rankings are similar,
the reason we do not cluster based on \psdist\ and \gdist\ is as follows. 

As we introduce in Section~\ref{subsec:scoring}, the mapping system works at real time.
That is, the paths need to be monitored 
a few times in a minute. However, the task of clustering itself should be run much less frequently.
Therefore, the clusters should be stable over time. That is, the clients should not migrate from
one cluster to the other one frequently, at least until the next run of cluster generation. 
For that reason, the clusters should be formed based on a more stable metric than latency. 
We know that latency is prone to fluctuations due to many factors, such as queueing 
time, server response time etc. Moreover, in order to cluster based on \psdist\ and
\gdist\ we still need to know the latency between server regions and the clients which
is the challenge that we tackle to solve in the first place. Therefore, we propose a clustering
method that is based on inter domain routing which is more prone to frequent to fluctuations
compared to latency on the paths.

Finally, in addition to \psdist\ and \gdist\ , one can define a metric that considers the
degree of the latencies instead of their orders. That is, one can categorize latency values 
as (really low, low, ok, high, really high) by defining a lower and upper boundary latency values
for each category and then measure the similarity between categorical vectors. For brevity, in this paper, 
we only note that such similarity metric yields similar results to \psdist\ and \gdist. 
We refer the reader to \cite{mytech} for details.

\section{Datasets}   \label{sec:dataset}  In this study, we use traceroute measurements and BGP announcements that are collected in the Akamai's CDN. 
We collected each of them on two separate days, July 2, 2014 (\dsjuly) and January 24, 2016 (\dsjan). 

\mpara{1. Traceroute Measurements.} We collected traceroute measurements from Akamai server regions to local DNS clients. 
For \dsjuly, the measurements are taken from 2211 server regions to 20110 clients. For \dsjan, the measurements are taken 
from 2073 server regions to 23004 clients. The Akamai server regions are spread across the globe. 
The DNS clients are located in six European countries (France, Germany, Spain, Italy, Switzerland, Belgium) 
and they belong to various ISPs. 

In practice, there are limitations on the number of times a DNS client can be tracerouted at a given time period. Such limitations are set by
the ISPs in order not to keep DNS clients busy. Therefore, each DNS client is tracerouted from a subset of the Akamai server regions. 
The number of server regions (known vector entries) per DNS client is 20 at minimum. Therefore we set $k = 20$ for \psdist. 

Each measurement from a server region to a DNS client consists of 
three consecutive ICMP packets and among these three, we use the one with the minimum latency.
Using these latency values we generate a rank vector for each DNS client. 

\mpara{2. BGP Announcements.} We use a collection of BGP tables collected from Akamai routers. 
For \dsjuly, the tables are collected from 233 peer routers and consist of over 1.7M BGP paths to 
over 37K prefixes located in the six European countries (France, Germany, Spain, Italy, Switzerland, Belgium).
Using this dataset, we map our DNS client IPs to their longest matching BGP prefixes. 20110 DNS servers in \dsjuly\
map to 5491 unique prefixes.  

For \dsjan, the tables are collected from 297 peer routers and consist of over 790K BGP paths to over 48K prefixes 
located in our six European countries. 
Using this dataset, we map our DNS client IPs to their longest matching BGP prefixes. 23004 DNS clients
map to 3272 unique prefixes.



\section{BGP Prefixes as Pre-Clusters}  \label{sec:pfx-eval}  
\begin{figure*}[tbp]
\centerline{
\includegraphics[width=0.33\textwidth]{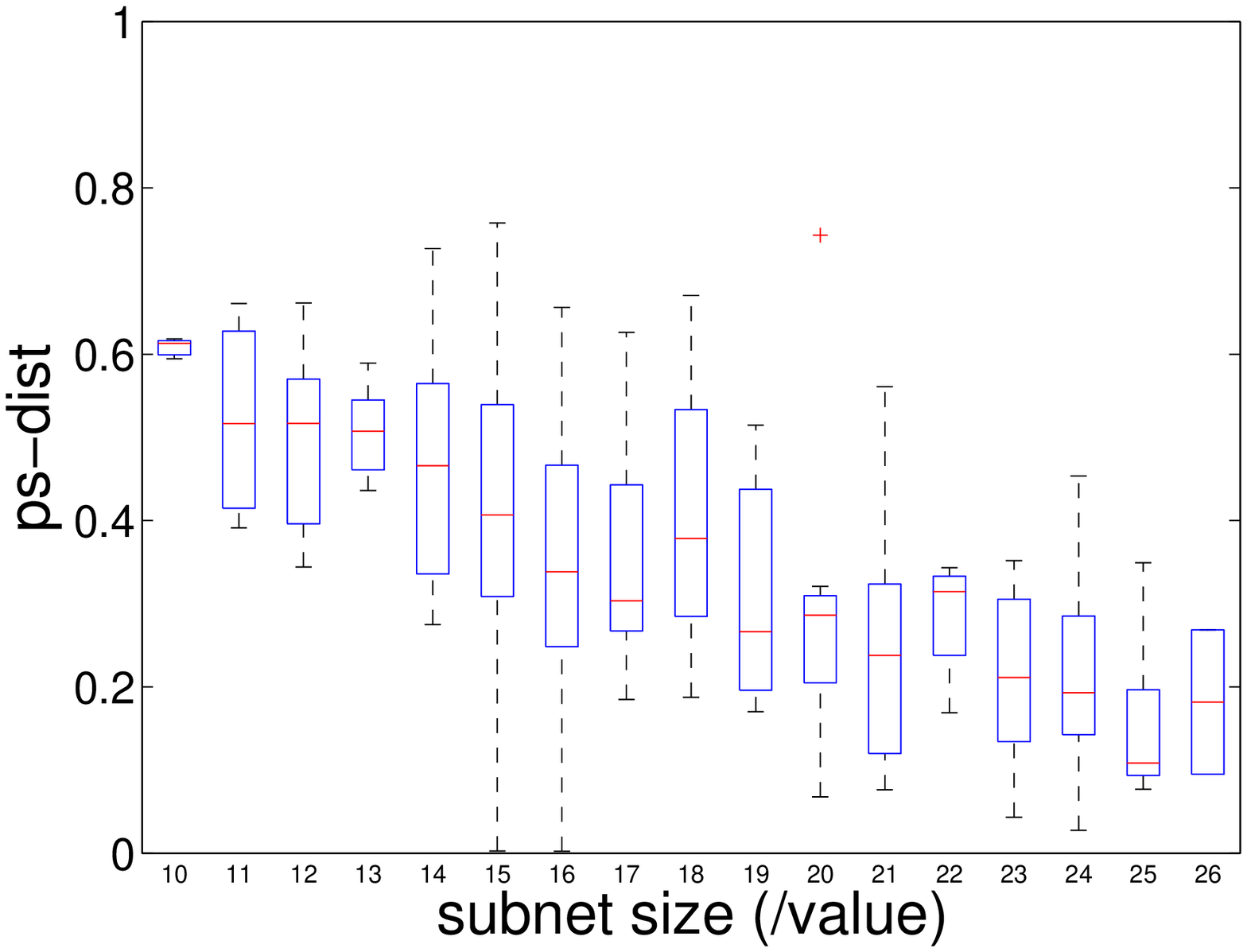}
\includegraphics[width=0.33\textwidth]{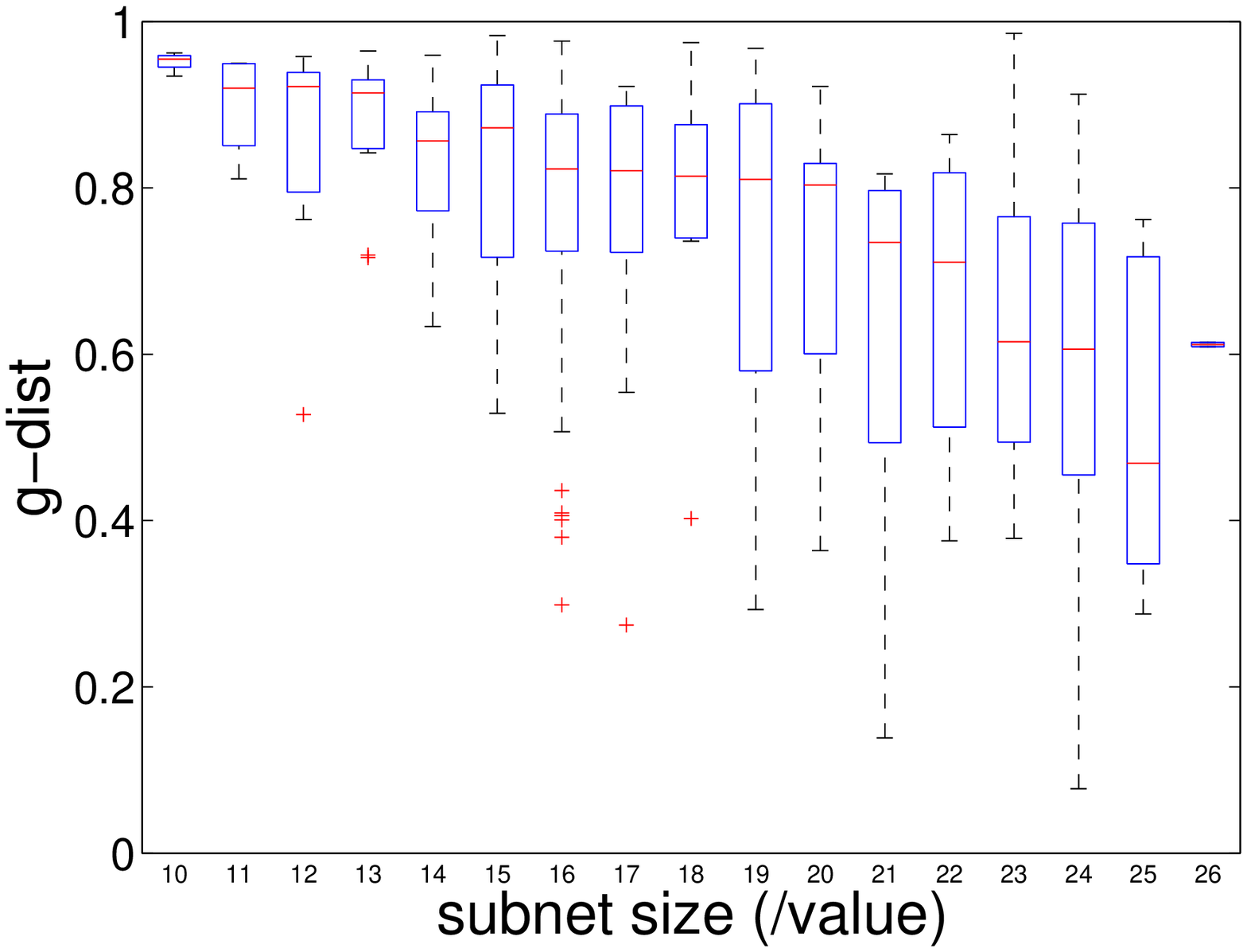}
\includegraphics[width=0.32\textwidth]{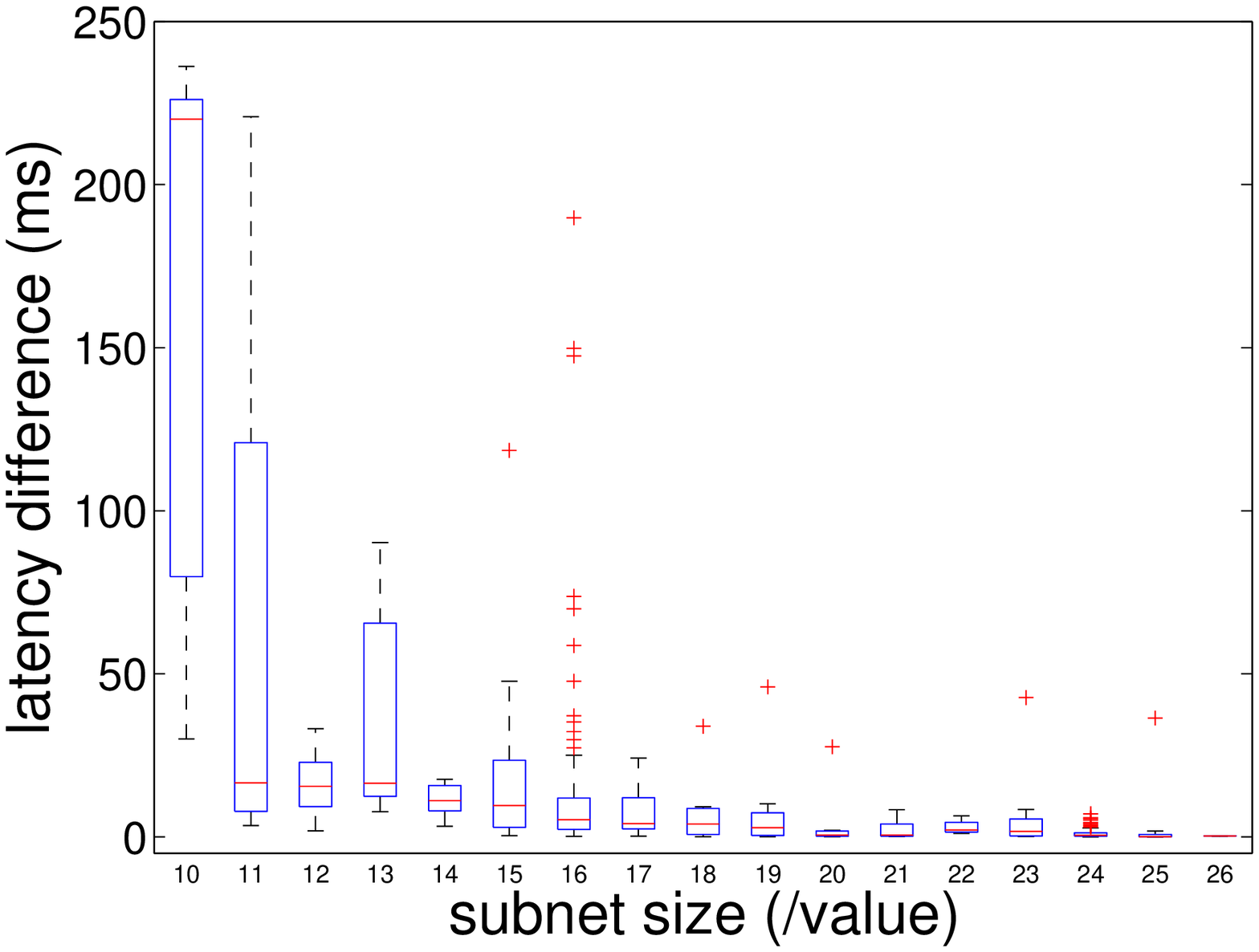}
}
\centerline{
\includegraphics[width=0.33\textwidth]{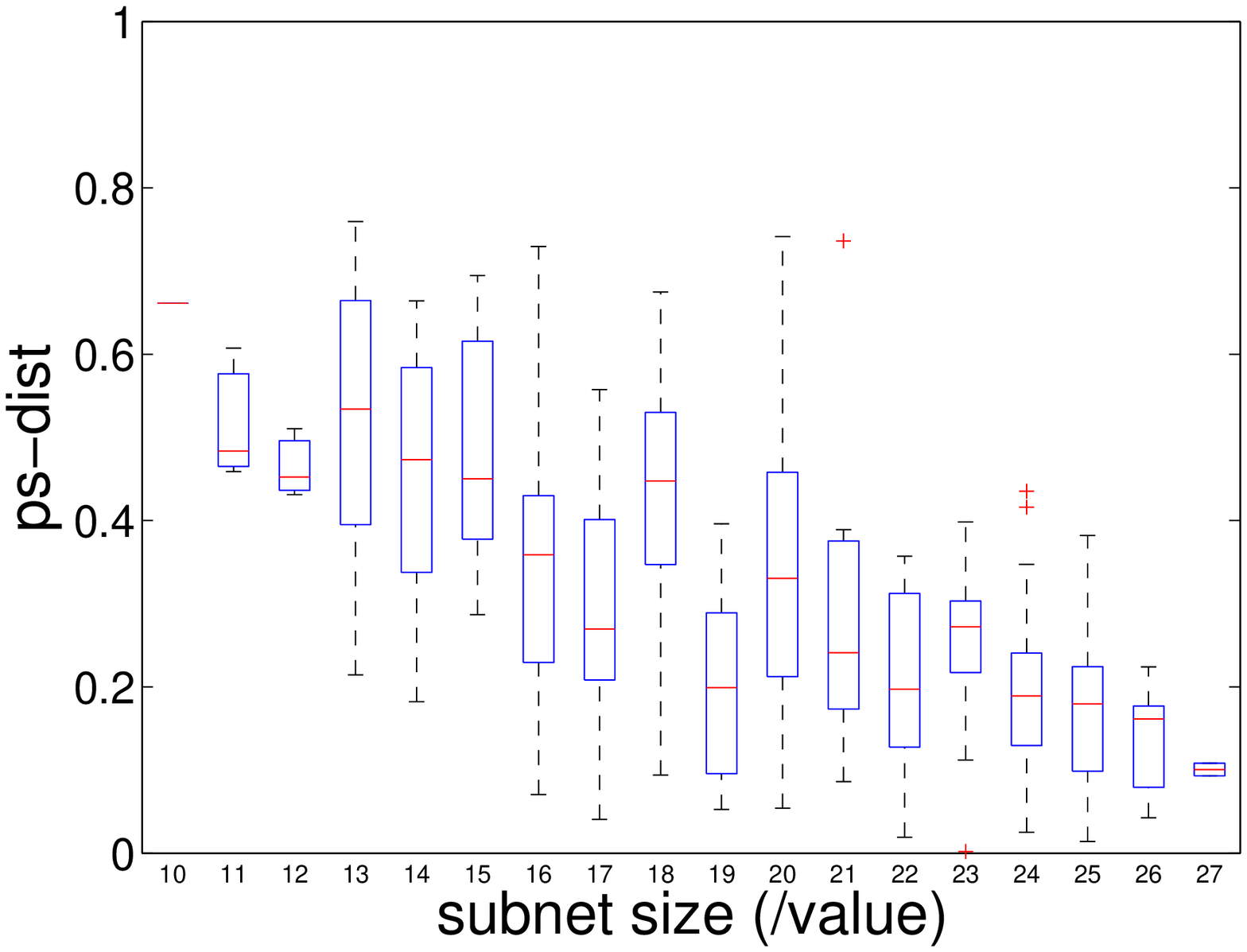}
\includegraphics[width=0.33\textwidth]{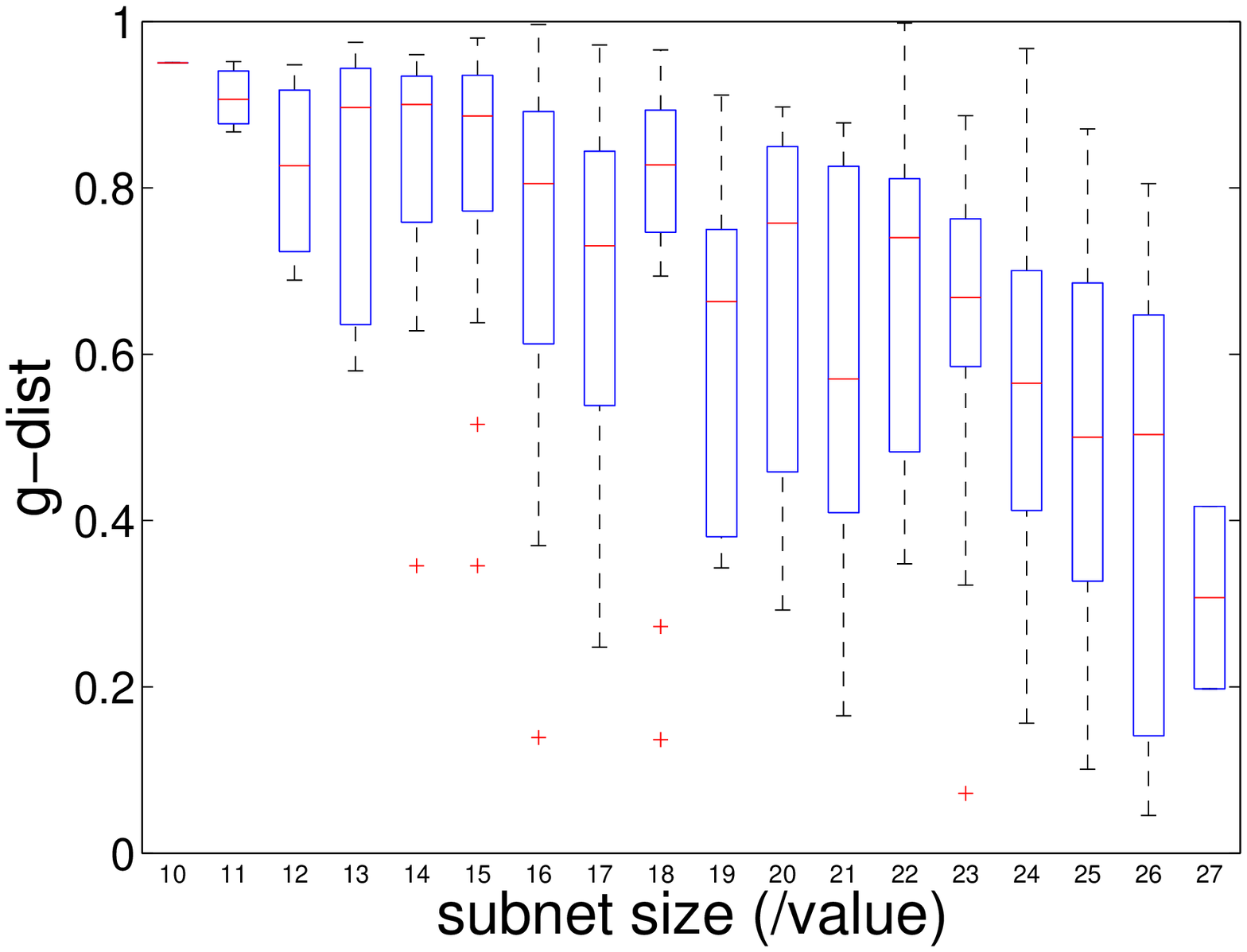}
\includegraphics[width=0.32\textwidth]{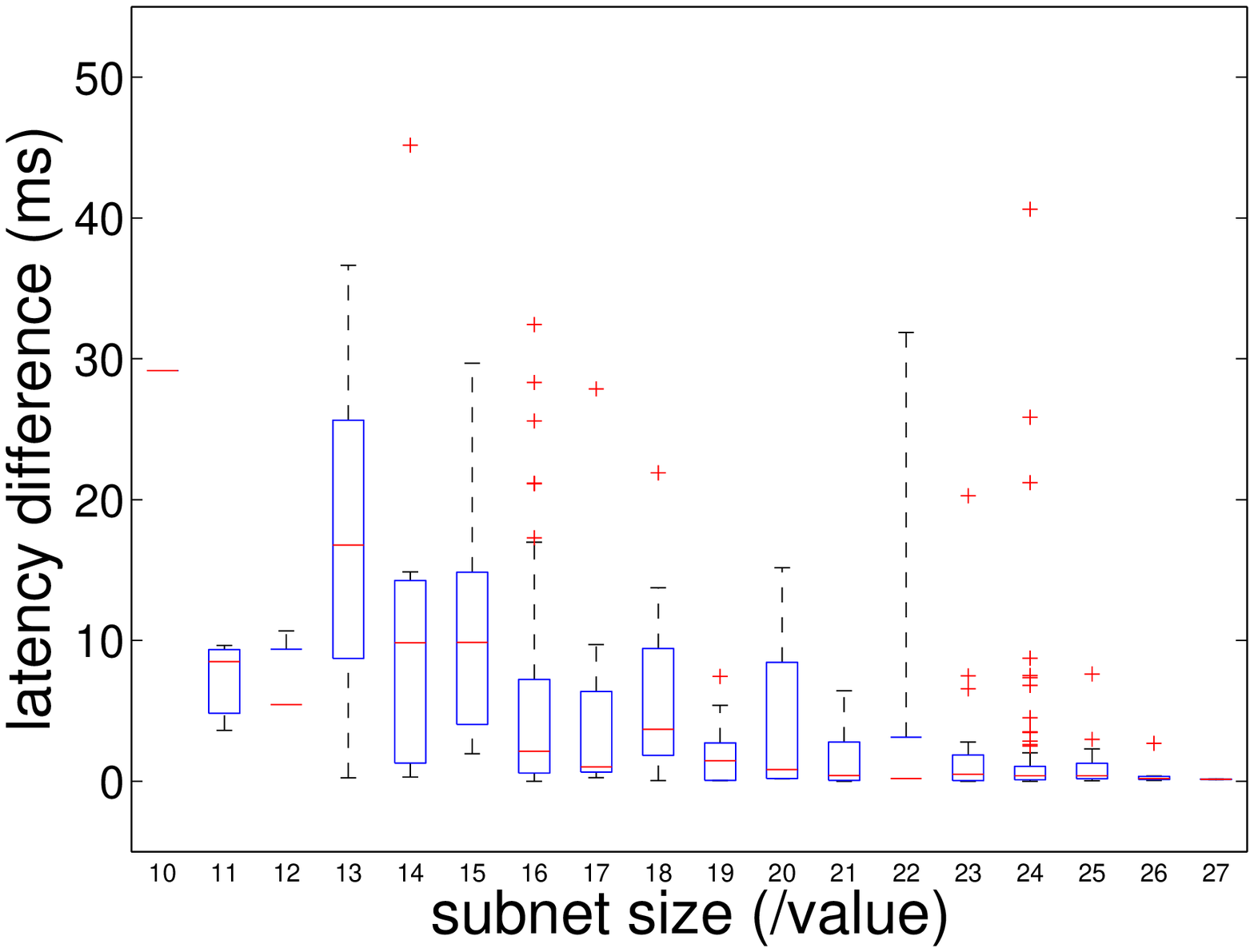}
}
  \centerline{\hfill\hfill(a)\hfill\hfill\hfill(b)\hfill\hfill(c)\hfill\hfill}
\caption{\psdist\ (a), \gdist\ (b),  latency difference (c) vs. BGP prefix length. In each box, the red line is the median, 
the upper and the lower end of the box are the first and third quantiles, respectively.
The upper and lower ends of the whiskers are the maximum and minimum values in 
the data, respectively. The individual red points are outliers. The top row shows the results from \dsjuly\ and the bottom 
row shows the results from \dsjan.} 
\label{fig:dist_vs_pfxLength}
\end{figure*}

One can group clients by their longest matching network prefixes as advertised by the BGP system. 
Such grouping is atomic in terms of routing because BGP dictates that all clients matching a 
destination prefix are routed the same way at the inter-domain level. 

Our intent is pre-grouping the clients by their prefixes 
\footnote{Throughout this paper, we will refer \emph{longest matching BGP prefix} as \emph{prefix}.}
and using these prefixes as data points for clustering instead of using individual client IPs. 
Our aim is providing a faster and more efficient partitioning 
by reducing the number of data points to be clustered.
 
To that end, we ask whether
pre-grouping clients by their prefixes yields good partitioning. 
That is, we test whether two clients from the same prefix are
close to each other in terms of their server rankings. 
In order to do this, we first define \emph{optimal partitioning} to set the benchmark 
for testing goodness of prefix clustering. 

\mpara{Optimal Partitioning.} Let $X$ be the set of clients and 
${\cal P}^r = \{ {\cal P}_1^r, {\cal P}_2^r, \dots, {\cal P}_n^r\} $ 
be a partitioning on $X$ such that every $X_i \in X$ belongs to one partition ${\cal P}_l^r$.
Let $x_1$ and $x_2$ be the ranking vectors of $X_i$ and $X_j$, respectively. 
If two clients, $X_i$ and $X_j$, prefer the same servers in the same order as their top-$r$
choices then they belong to the same partition, i.e.\ $X_i \in {\cal P}_l^r$ and $X_j \in {\cal P}_l^r$.

For instance, in the example in Figure~\ref{tab:example}, for ${\cal P}^2$, $X_5$ and $X_6$
are in the same partition since they prefer $T_1$, $T_2$ in the same order as their top-2, 
whereas none of the other clients will be partitioned together as they all have different top-2
orderings.  

The reason we call such partitioning \emph{r-optimal} is that it does not allow two clients
to be in the same partition if they don't agree on the same servers in the exact same order 
for their top-$r$ choices. Therefore it sets the tightest constraints and guarantees the 
most compact clusters for the particular choice of $r$.

Also notice that, by definition, optimal partitioning groups clients based on their \gdist. For instance,
in a \emph{1-optimal} partition, \gdist\ between any pair of client is not greater than 0.5. Similarly,
in a \emph{r-optimal} partition for a slightly large value of $r$, \gdist\ between any pair of client is close 
to 0.  

\begin{figure}[tbp]
\centerline{\includegraphics[width=0.45\textwidth]{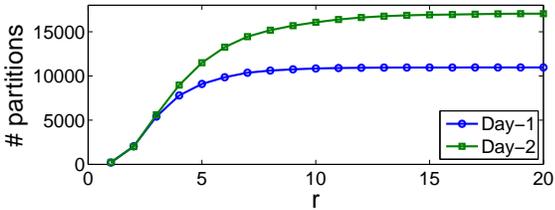}}
\caption{The number of optimal partitions for a range of $r$ values.} 
\vspace{-1.0em}
\label{fig:base_numParts}
\end{figure}

We apply optimal-partitioning on the set of 20110 (\dsjuly) and 23004 (\dsjan) 
rank vectors that we described in Section~\ref{sec:dataset}.
Figure~\ref{fig:base_numParts} shows the number of partitions for $r$ values from 1 to 20. 
The number of partitions increases as $r$ increases. 
This is expected since for large values of $r$, there exists larger number of combinations
for top-$r$ rankings. For $r$ values of 1,2,3, the number of partitions are around 200, 2000, 5400 respectively. 
For $r$ values greater than 6, the number of partitions converges to a number slightly greater than 10000 (\dsjuly) and
14000 (\dsjan). That is, on average, there exists only one or two clients under the same partition for $r$ values greater than 6.

Next, for each partition ${\cal P}_l^r$, let $t_l^r$ be the number of clients in the partition
and  $\mu_l^r$ be the mean of all pairwise \psdist\  values within the partition as written below:

\begin{equation}
	\mu_l^r = \frac{\sum_{X_i, X_j: X_i, X_j \in {\cal P}_l^r} \psdist(x_i,x_j)}{t_l^r * (t_l^r-1)}
\end{equation}

Then $M^r =  \{ \mu_1^r , \mu_2^r, \dots, \mu_n^r \} $ is the distribution of mean of pairwise distances for 
the partitioning ${\cal P}^r$. 

\mpara{Comparing BGP prefix clusters with optimal partitioning.} One variable in BGP prefixes is the 
length, i.e. the number bits in the subnet mask. The longer a prefix, the less 
clients it has and the more likely that the clients are close to each other in the network. 

In order to understand the role of the prefix length, first we cluster clients by 
their longest matching prefixes as found in the BGP dataset we described 
in Section~\ref{sec:dataset}. Note that each client is matched with only one prefix. 
Second, we compute the pairwise \gdist\ and \psdist\ between each 
and every pair of clients that are from the same prefix. Then we compute the
average of pairwise distances within each prefix. 
Third, we group these averages by the length of their prefixes.
In our dataset, the length of BGP prefixes vary from /10 to /26 in \dsjuly\ and
/10 to /27 in \dsjan. 

Figure~\ref{fig:dist_vs_pfxLength} shows the statistics of each prefix length group as a separate box. 
The top row shows the results from \dsjuly\ and the bottom row shows the results from \dsjan.
In Figure~\ref{fig:dist_vs_pfxLength} (a) we see that there are client pairs from small prefixes 
(e.g. /19- /26) that are close to \emph{1-optimal}. However, in large prefixes (e.g. /10, /11) the clients are 
far away from each other. In Figure~\ref{fig:dist_vs_pfxLength} (b), we see the same trend, i.e. 
as the lengths of the prefixes increase, the distances within the prefix group decrease. 
This suggests that smaller subnets generate more compact clusters. 

Next we investigate how large the latency can get for a client due to clustering. 
We compute the worst possible latency difference for each client in a given prefix as follows.
For each client $X_i$ in a given prefix, we find the largest latency server with respect to $X_i$, 
say $T_j$ s.t. $T_j$ is the top-1 for some other client in the same prefix. 
Then, we compute how much the latency to $X_i$ increases if $T_j$ is assigned to $X_i$ instead of 
its top-1 server. Figure~\ref{fig:dist_vs_pfxLength} (c) shows the average of such latency differences per prefix grouped by 
prefix length. The figure shows that the latency difference drops significantly as the prefix length grows 
and the difference is around 0 for small prefixes. 

\begin{figure}[tbp]
\centerline{\includegraphics[width=0.42\textwidth]{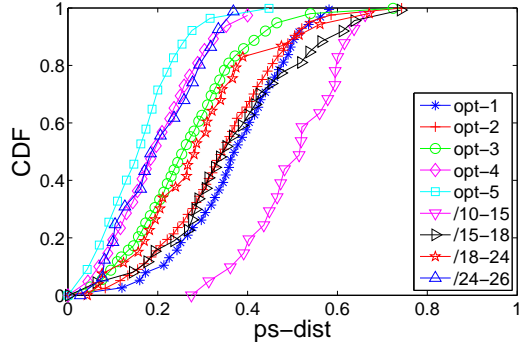}}
\caption{\dsjuly: Distributions of optimal partitioning for $r$ values from 1 to 5 and avg. pairwise \psdist\ within each prefix-length 
group.} 
\vspace{-0.9em}
\label{fig:basecase}
\end{figure}

\begin{figure}[tbp]
\centerline{\includegraphics[width=0.42\textwidth]{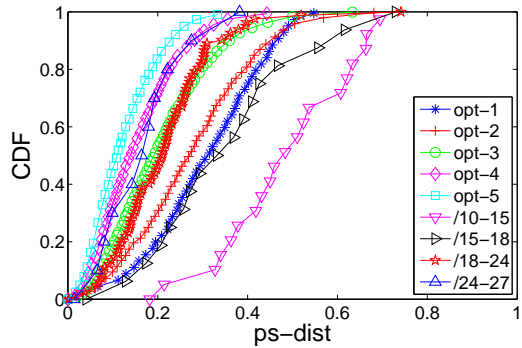}}
\caption{\dsjan: Distributions of optimal partitioning for $r$ values from 1 to 5 and avg. pairwise \psdist\ within each prefix-length 
group.} 
\vspace{-1.4em}
\label{fig:basecase2}
\end{figure}

Next, we test grouping the clients by prefix against the optimal partitioning 
with a range of $r$ values. We compute the average pairwise \psdist\ 
within each prefix (as described above). Then we divide the set of these average values 
into four groups by the length of their prefixes, 
/10-/15, /15-/18, /18-/24, /24-/26. 
We compare the distribution of values in each group with 
$M^r$ for $r = 1 \dots 5$ in Figure~\ref{fig:basecase} and Figure~\ref{fig:basecase2}
for \dsjuly\ and \dsjan, respectively. 
We see that the distributions of the prefixes from the /15-/18, /18-/24, and /24-/26
groups are very close the \emph{2-optimal}, \emph{3-optimal}, and \emph{4-optimal}, 
respectively. 
To quantify the results, we run Kolmogorov-Smirnov tests to check the similarity of 
distributions from prefixes and their corresponding optimal partitions. The /18-/24 and /24-/26
groups passed the test at the 1\% significance level. In addition, we note that 20110
clients in \dsjuly\ map to 5491 prefixes and 23004 clients in \dsjan\ map to 3272 prefixes.
That is, the mapping to BGP prefixes reduce the set of clients almost as much as 
\emph{3-optimal} partitioning and better.  

\begin{figure*}[tbp]
\centerline{
\includegraphics[width=0.25\textwidth]{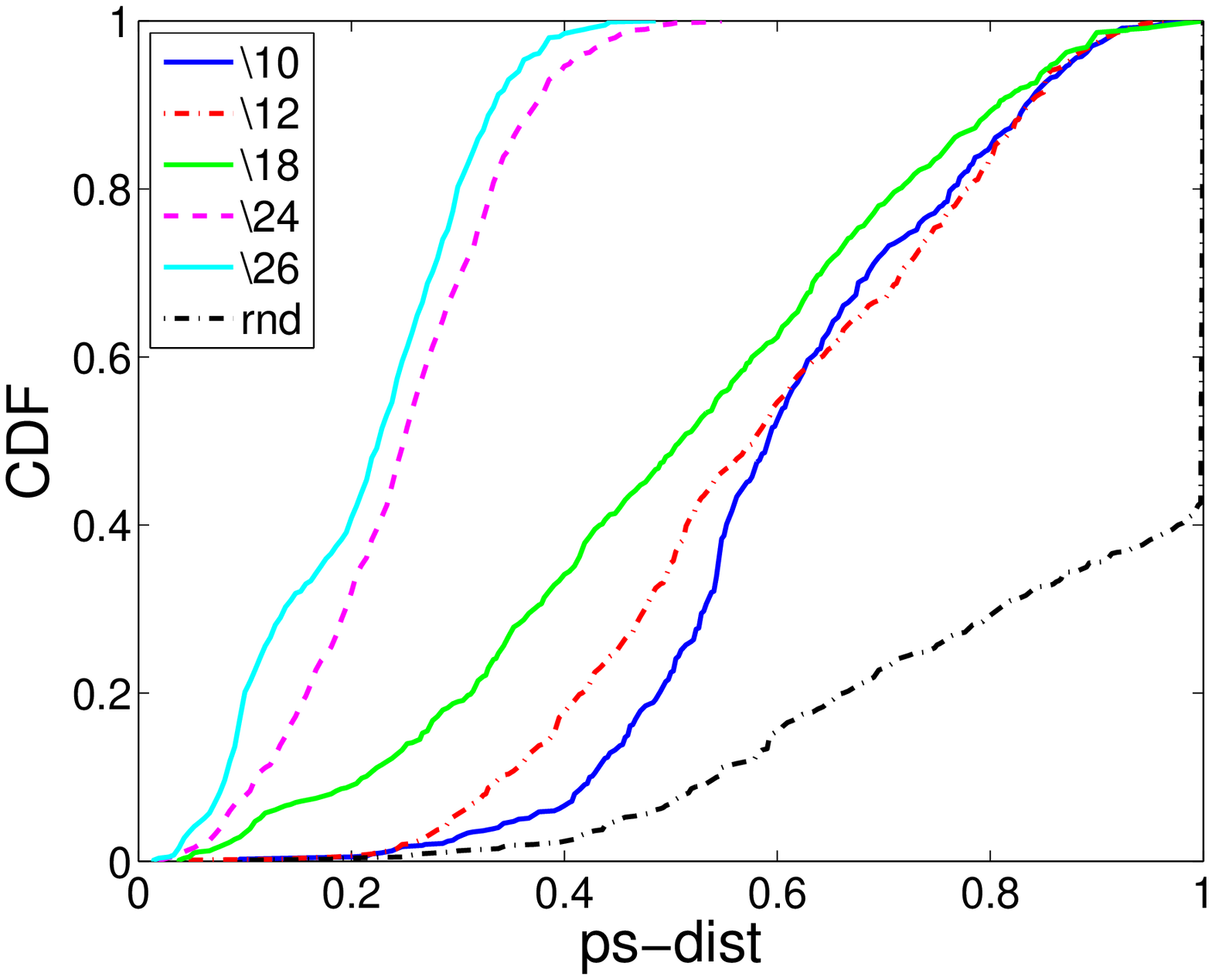}
\includegraphics[width=0.25\textwidth]{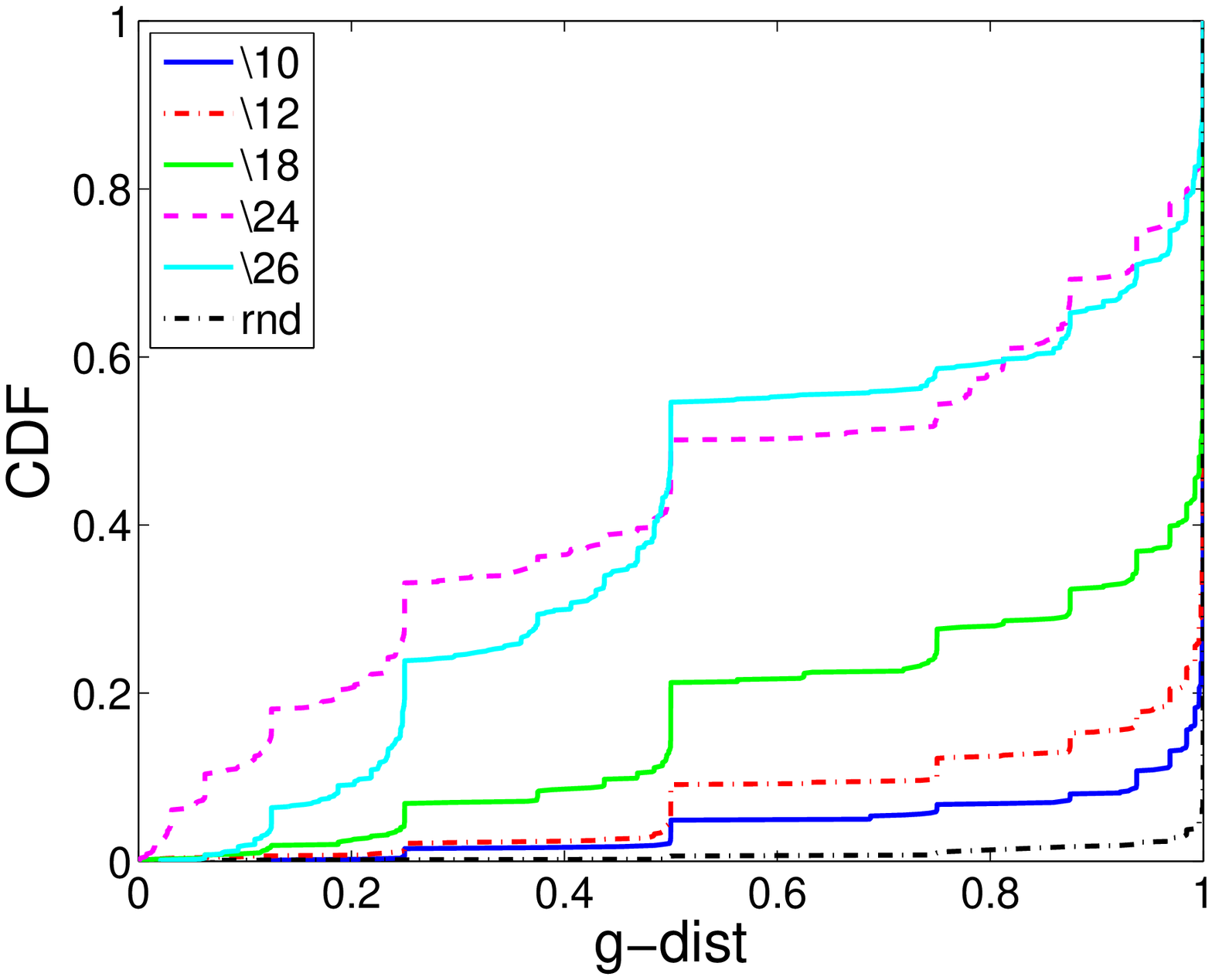}
\includegraphics[width=0.25\textwidth]{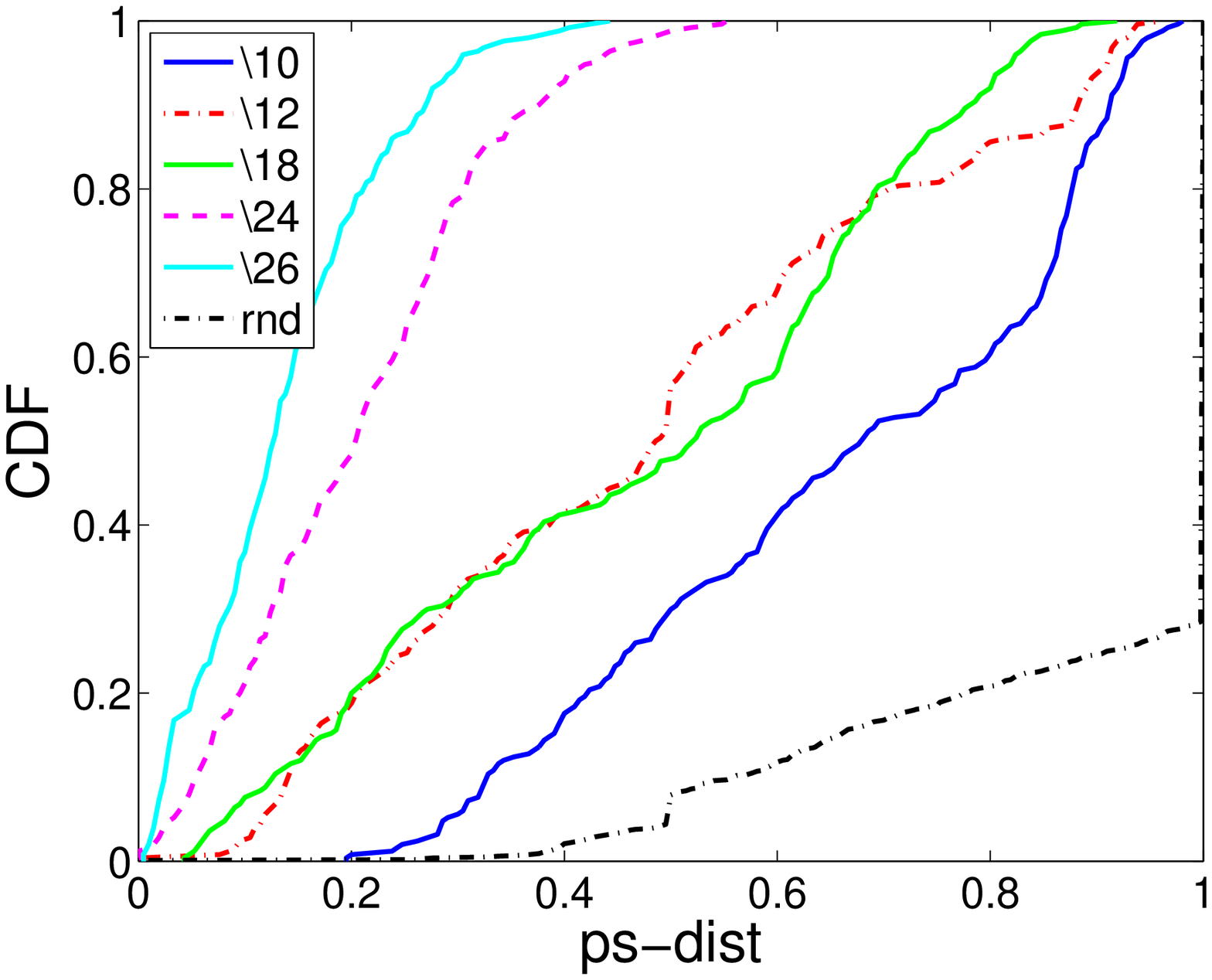}
\includegraphics[width=0.25\textwidth]{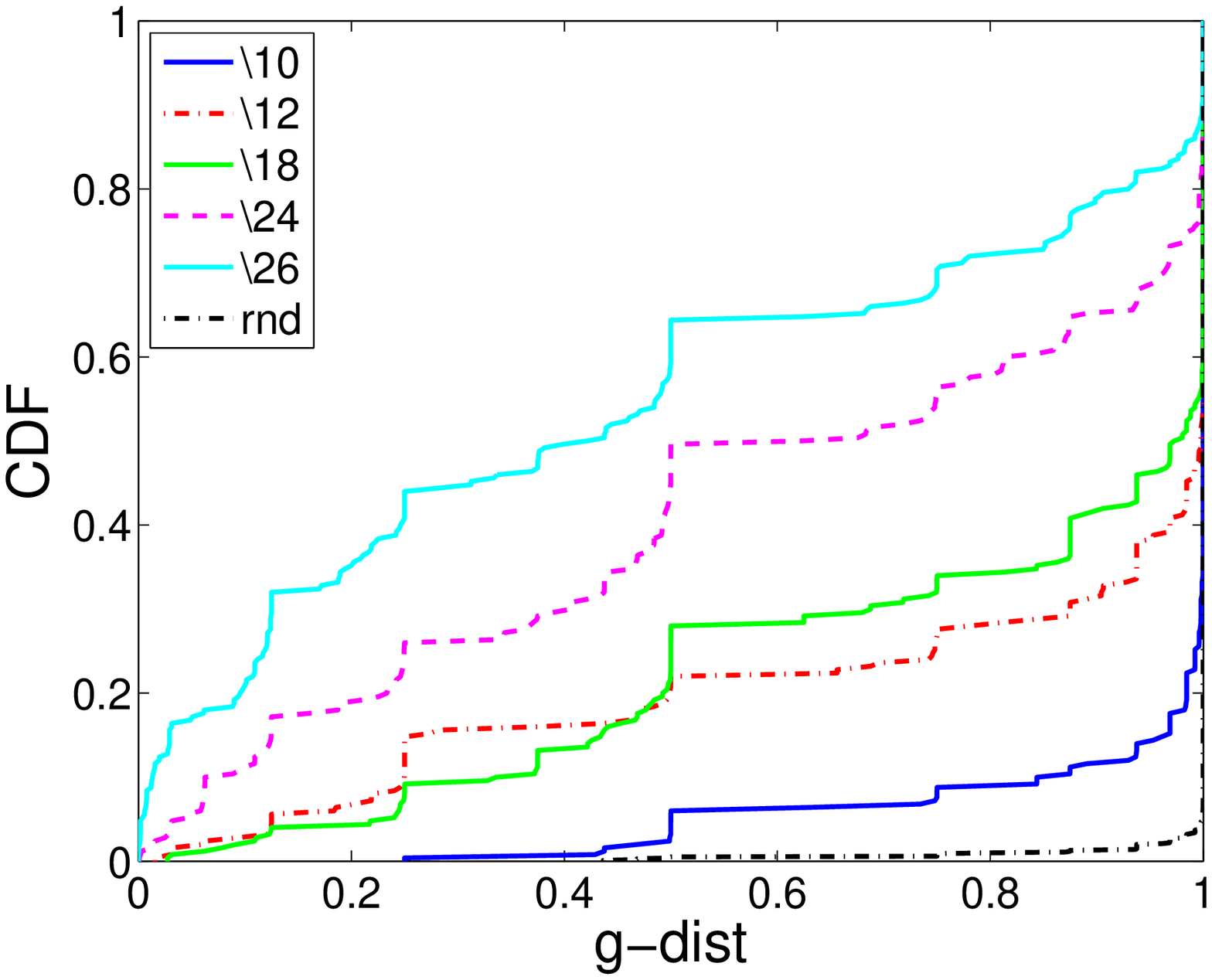}}
  \centerline{\hfill\hfill(a) \dsjuly\ \hfill\hfill\hfill(b) \dsjan\ \hfill\hfill}
\caption{Distribution of pairwise distance values between 1000 pairs of clients that are from 
the same BGP prefix and 1000 pairs of clients that are from randomly selected, different BGP prefixes.} 
\label{fig:bgp_random_pfx}
\end{figure*}

Finally, we compare the pairwise \psdist\ and \gdist\ between clients that are from the
same prefixes with the ones from different prefixes. We randomly sample 1000 client pairs 
that are from the  same  prefix and 1000 client pairs that are from different prefixes.
Figure~\ref{fig:bgp_random_pfx} shows the distribution of their pairwise distances. 
We see that for both metrics, the distance values between clients from the same prefixes 
are much lower compared to the the distance values between clients from different prefixes.

We conclude that grouping clients by their longest matching BGP prefixes 
generates successful clusters. Moreover, the similarity of server ranking within
prefixes (especially the small ones) are close to the optimal. 
For that reason, one can use prefixes as data units for further 
clustering. We suggest dividing prefixes of small subnet lengths
(e.g.\ /10- /12) into smaller subnets (e.g.\ /24s) in practice.



 
\section{Routing-Aware Clustering}  \label{sec:clust-eval}  One factor that has large impact on the latency between two ends is the routing path between them.
In this section, we apply a routing-aware clustering on the set of prefixes and study
these clusters for the server ranking problem.

\subsection{Correlation Between Routing State and Server Ranking}
\label{subsec:rs}

Our hypothesis is that if the routing paths to two destination prefixes, 
$p_1$ and $p_2$ are similar from a set of server regions, then $p_1$ and $p_2$ 
experience similar latencies from these server regions and therefore rank them similarly. To test this hypothesis, 
first we revisit the notion of \emph{routing similarity} as introduced in \cite{Gursun12}. 

Let $A$ be the set of all ASes in the Internet s.t. $A = \{a_1, a_2, \dots, a_t\}$ and $p_1$ and $p_2$ are 
announced by $a_1$ and $a_2$, respectively. We assume that for any $a_i$ there is a unique\footnote{This assumption is relaxed later.} 
$a_j$ which is the next hop AS on the path to $p_1$. That is, $nexthop(a_i, p_1)  = a_j$, and  
$nexthop(a_1, p_1) = a_1$.
Then, the \emph{routing state} of $p_1$ is the collection of next hop choices to $p_1$ as described below: 
\
\begin{eqnarray*}
	\lefteqn{\mbox{\em routestate}(p_1) =}\\ && \langle\mbox{\em nexthop}(a_1, p_1), \mbox{\em nexthop}(a_2, p_1), ..., \mbox{\em nexthop}(a_t, p_1)\rangle
\end{eqnarray*}

Given \emph{routestate}$(p_1)$ and \emph{routestate}$(p_2)$, we measure the routing similarity 
between $p_1$ and $p_2$ by the number of ASes that prefer the same next hops to $p_1$ and $p_2$ 
as defined below:

\[ 
\mbox{\em rsim}(p_1, p_2) = \#\{a_i \,|\, \mbox{\em nexthop}(a_i, p_1) = \mbox{\em nexthop}(a_i, p_2)\}
\]

Similarly, the routing dissimilarity between $p_1$ and $p_2$ is called \emph{Routing State Distance} (\rsd) 
and it is defined below:

\[ 
\mbox{\em rsd}(p_1, p_2) = \#\{a_i \,|\, \mbox{\em nexthop}(a_i, p_1) \neq \mbox{\em nexthop}(a_i, p_2)\}
\]

\mpara{Applying \rsd\ on the BGP dataset.}
Having defined routing similarity, we seek to compute \rsd\ for prefix pairs in our BGP dataset. 
However, there are two main issues. First, for a given prefix, we can not observe a \nexthop\ from every AS 
in the Internet. To address this case when \nexthop$(a_i, p_1)$ is not available, 
\rsd$(p_1,p_2)$ is approximated by the fraction of \emph{known} 
next hops in which $\routestate(p_1)$ and $\routestate(p_2)$ differ, times the total 
number of ASes in set $A$.  This normalizes \rsd\ so that it always
ranges between zero and the total number of ASes in set $A$, i.e.\ $|A|$.
We called the normalized version \RSD. 

The second issue is that for some AS-prefix pairs\\
\nexthop\ function is not uniquely defined, 
that is traffic destined for the same prefix may take different next hops
e.g. when an AS uses hot-potato routing. We address this problem 
as in \cite{Gursun-thesis}, i.e.\ by dividing each AS in set $A$ 
into a minimal set of sub-ASes such that for each (sub-AS, prefix) 
pair there is a unique next hop AS for the prefix. We call this extended
version of $A$, $A'$.  For the dataset of \dsjuly, $|A'| = 1460$, and for \dsjan, 
$|A'| = 1404$. There are around 50 next hops in average per prefix in both days. 
Note that both implementation considerations are
discussed in \cite{Gursun-thesis} in great detail. 

Having addressed these two issues, we compute, \rsim\ and \RSD\ for
each pair of prefixes in our BGP dataset.
Notice that due to the normalization, unknown next hop, and multiple next hop 
issues described above, 
\RSD$(p_1,p_2)$ is not simply $|A| - $\rsim$(p_1, p_2)$. 

Next, we seek to understand the correlation between routing similarity and 
server ranking of two prefixes. For each prefix pair we compute the minimum, 
average, and maximum \psdist\ and \gdist\ between the clients of the prefixes as defined below. 
Note that the \gdist\ counterparts are defined likewise. 

\begin{equation}
    \psdist_{min}(p_1, p_2) =  \underset{{X_1 \in p_1, \atop X_2 \in p_2}}{\mbox{min}} \: \psdist (x_1,x_2)
\end{equation}

\begin{equation}
     \psdist_{max}(p_1, p_2) = \underset{{X_1 \in p_1, \atop X_2 \in p_2}}{\mbox{max}} \:  \psdist (x_1,x_2)
\end{equation}

\begin{equation}
     \psdist_{avg}(p_1, p_2) = \frac{1}{|p_1| |p_2|} \sum_{X_1 \in p_1, \atop X_2 \in p_2} \psdist (x_1,x_2)
\end{equation}

Note that $\psdist_{max}(p_1, p_1)$, $\psdist_{min}(p_1, p_1)$  are the maximum and minimum \psdist\ 
between two clients \\
within $p_1$, and they can be written as $\psdist_{max}(p_1)$ and  $\psdist_{min}(p_1)$, respectively.  
Similarly,  $\psdist_{avg}(p_1)$ is the average pairwise distances of clients in $p_1$.  Note that 
\gdist\ counterparts are defined likewise.

\begin{figure}[tbp]
\centerline{
\includegraphics[width=0.35\textwidth]{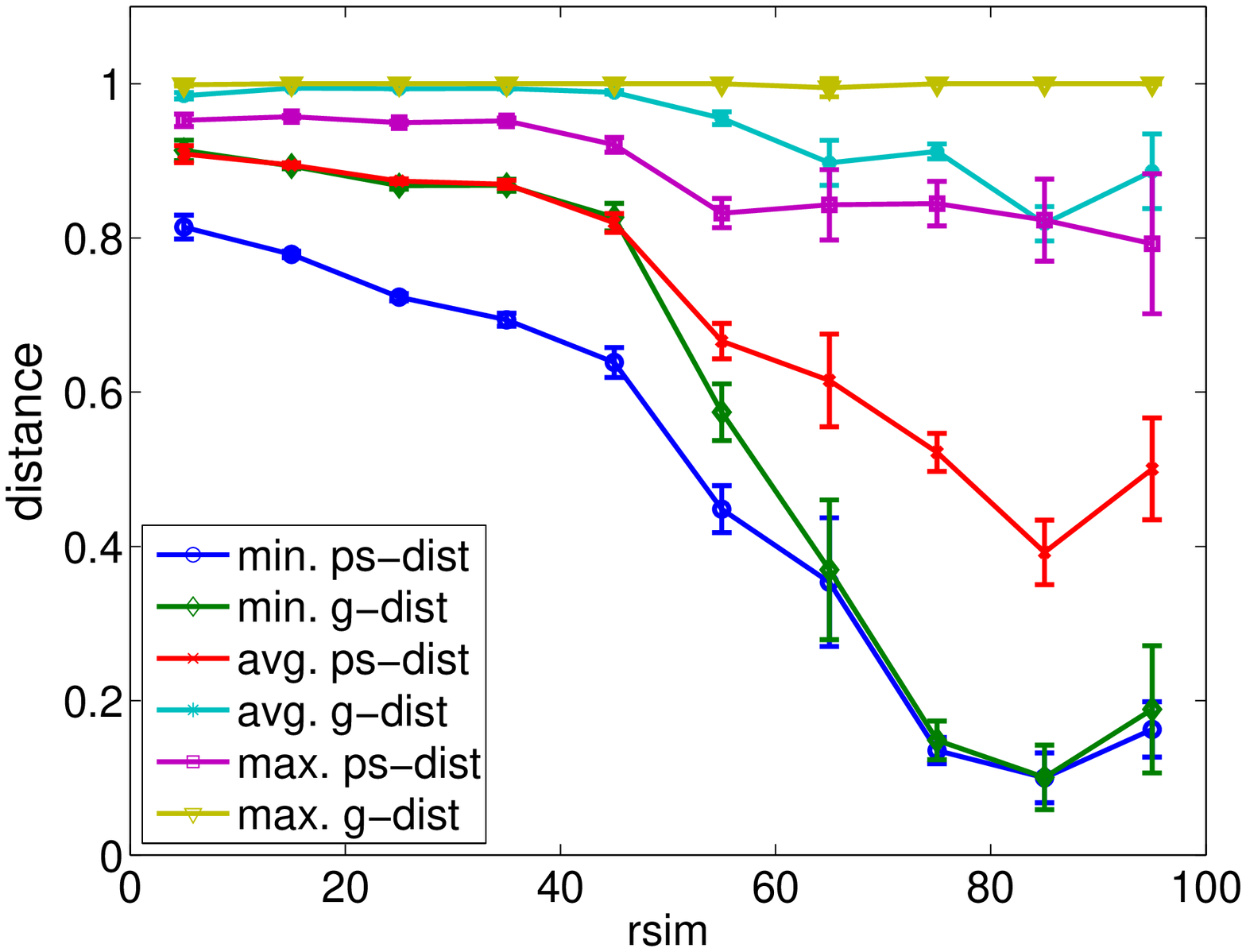}}
\caption{\dsjuly: \psdist\ and \gdist\ vs. \rsim\ between pairs of prefixes grouped by their \rsim\ values.} 
\label{fig:numNextHops}
\end{figure}

\begin{figure}[tbp]
\centerline{
\includegraphics[width=0.35\textwidth]{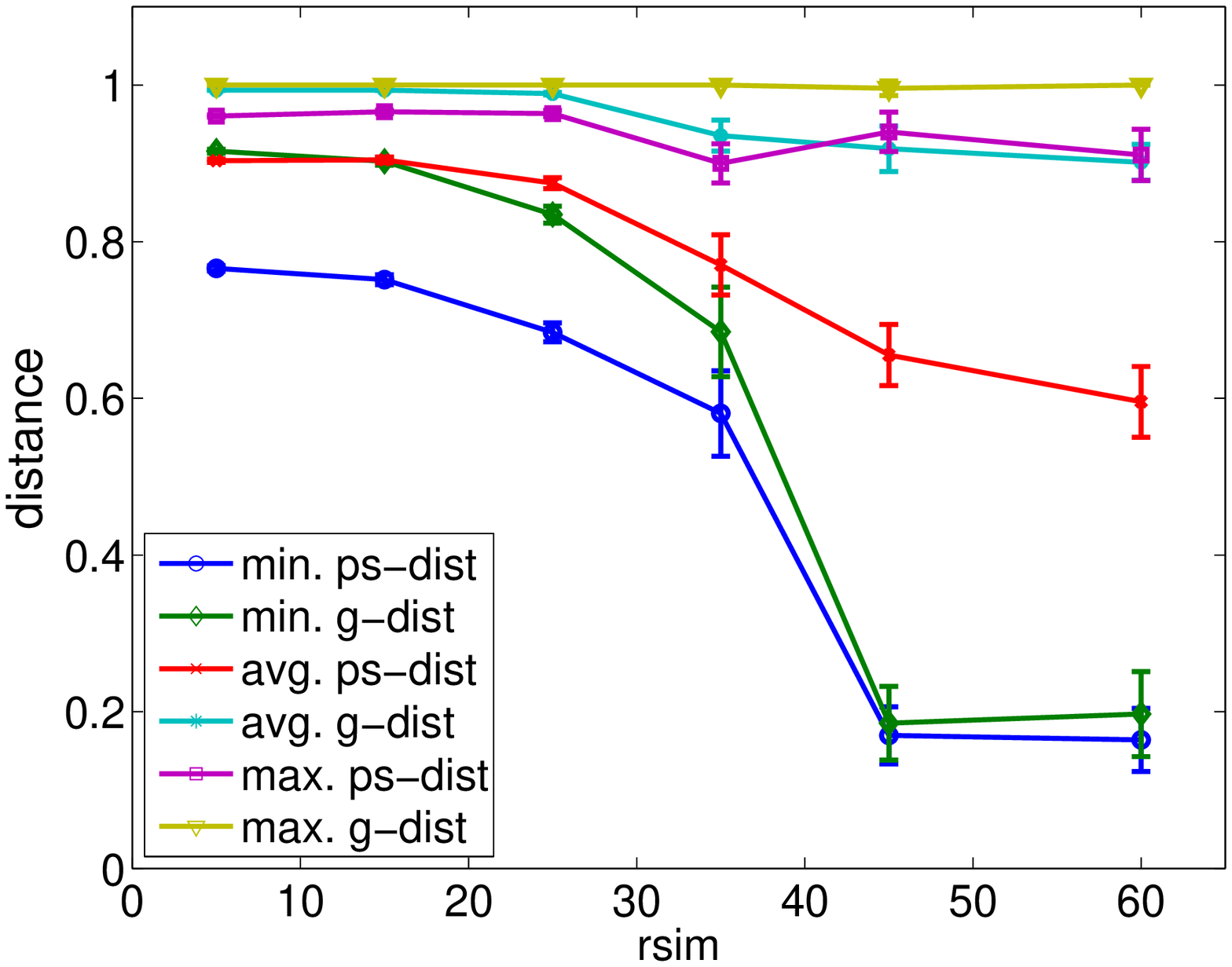}}
\caption{\dsjan: \psdist\ and \gdist\ vs. \rsim\ between pairs of prefixes grouped by their \rsim\ values.} 
\label{fig:numNextHops2}
\end{figure}

Given a pair of prefixes, Figure~\ref{fig:numNextHops} and Figure~\ref{fig:numNextHops2}
show the relationship between the number of ASes that prefer
the same next hops for these prefixes and their server ranking similarity. 
Prefix pairs are placed in buckets of size 10 according to their $\rsim$ values. Then, 
for each bucket, mean of the $\psdist_{min}$, $\psdist_{max}$, $\psdist_{avg}$, 
$\gdist_{min}$, $\gdist_{max}$, $\gdist_{avg}$
values are plotted with 95\% confidence interval. 

Both Figure~\ref{fig:numNextHops} and Figure~\ref{fig:numNextHops2} show 
that there is a strong correlation between the next hop choices for two prefixes
and their server ranking. As the number of ASes that choose the same next hops increases, the server ranking distance between two
prefixes decreases. The amount of decrease is larger for the $\psdist_{min}$ and $\gdist_{min}$ metrics, i.e.\ 
there is at least a pair of clients, one from each prefix, that are very close to each other if the similarity between prefixes is high.

\begin{figure}[tbp]
\centerline{
\includegraphics[width=0.35\textwidth]{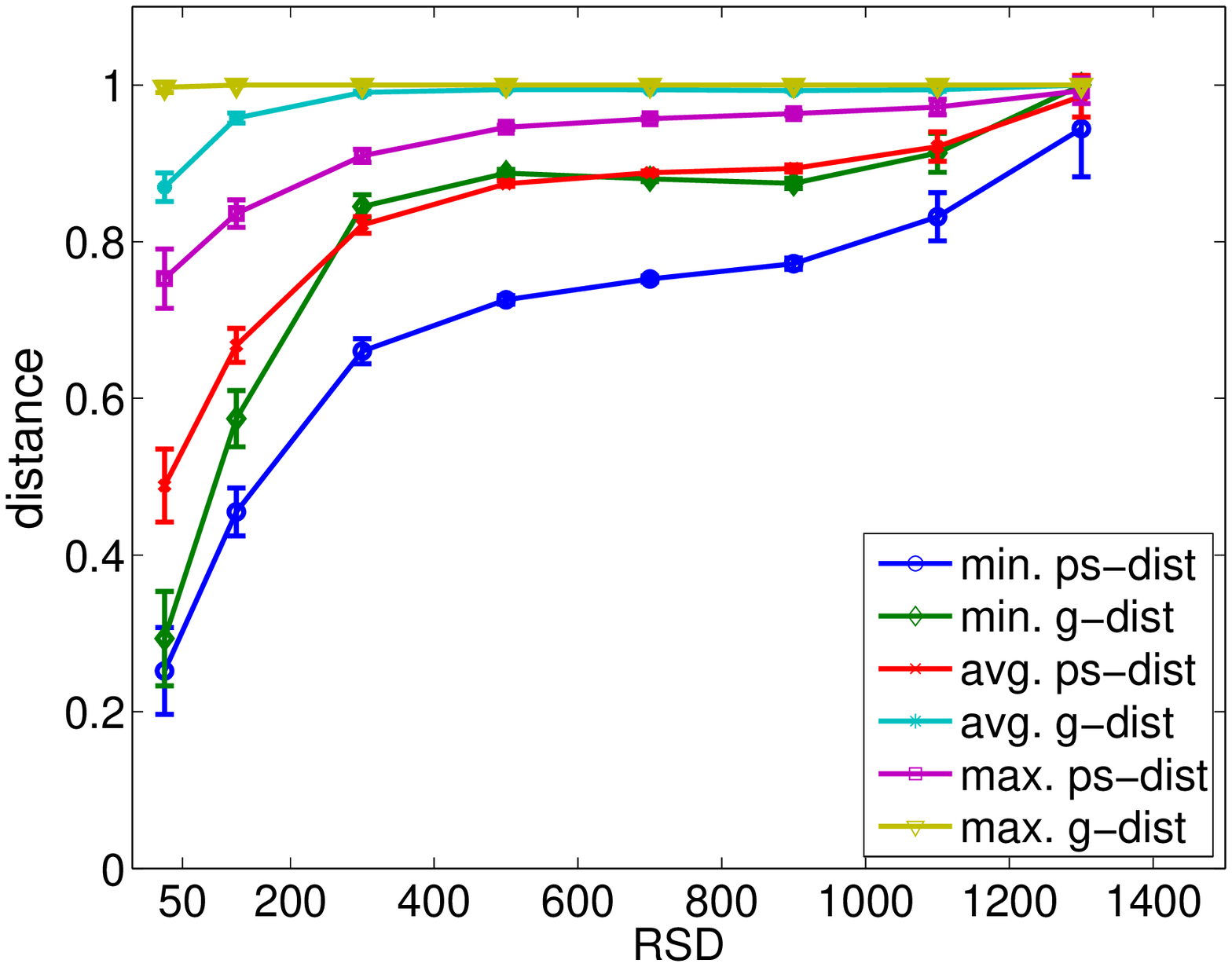}}
\caption{\dsjuly: \psdist\ and \gdist\ vs. \RSD\ between pairs of prefixes grouped by their \RSD\ values.} 
\label{fig:rsd_vs_dist}
\end{figure}

\begin{figure}[tbp]
\centerline{
\includegraphics[width=0.35\textwidth]{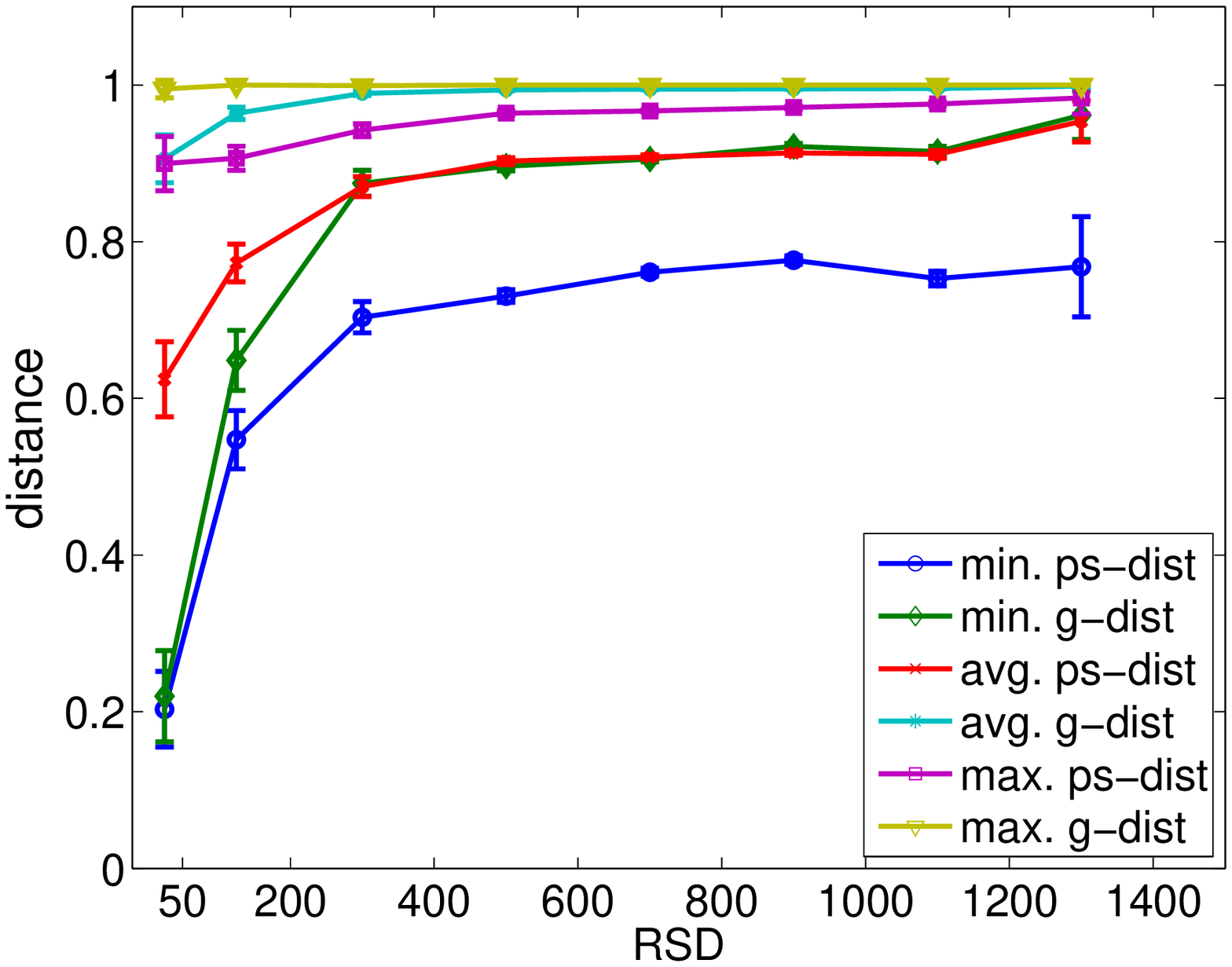}}
\caption{\dsjan: \psdist\ and \gdist\ vs. \RSD\ between pairs of prefixes grouped by their \RSD\ values.} 
\label{fig:rsd_vs_dist2}
\end{figure}

Similarly, in Figure~\ref{fig:rsd_vs_dist} and Figure~\ref{fig:rsd_vs_dist2} prefix pairs are placed in buckets of 
0-50, 50-200, 200-400, $\dots$ 1200-1400 according to their \RSD\ values. 
Then, for each bucket, mean of the 
$\psdist_{min}$, $\psdist_{max}$, $\psdist_{avg}$, $\gdist_{min}$, $\gdist_{max}$, $\gdist_{avg}$
values are plotted with 95\% confidence interval. Both figures show that 
as the \RSD\ between two prefixes decreases, their max, min, and average \psdist\ and \gdist\ decrease. 
In other words, two prefixes that are close to each other in the \RSD\ space are close to each other 
in \psdist\ and \gdist\ space too.

\subsection{Clustering by Routing Similarity}
\label{subsec:clust-rsd}

Having shown the correlation between routing and server ranking similarity, 
next we seek clustering prefixes in \RSD\ space. Intuitively, we are looking for 
a partitioning that minimizes the \RSD\ between two prefixes that are from the same cluster. 
One extended version of this intuition is formalized as \RSclustering\ problem 
in \cite{Gursun12} and solved by Algorithm~\ref{algo:pivot}.

\begin{algorithm}
\caption{The {\pivot} algorithm \label{algo:pivot}.}
\begin{algorithmic}[1]
\Statex A set of prefixes $P=\{p_1,\ldots ,p_{n}\}$  and a threshold $\tau\in [0,t']$.
\Statex {A partition $\calP$ of the prefixes}
\State pick a random prefix $p\in P$
\State create a cluster $C_p=\{p'\mid \RSD(p,p')\leq \tau\}$
\State $P = P\setminus C_p$
\State {\pivot}$(P,\tau)$
\end{algorithmic}
\end{algorithm}

The inputs of the algorithm are the set of prefixes, their pairwise {\RSD} values,
and a threshold parameter $\tau\in [0,t']$, where $t' = |A'|$
is the maximum possible value of \RSD. The algorithm works as
follows: Starting from a random prefix $p$, it finds all
prefixes that are within the distance $\tau$ from $p$. All these prefixes
are assigned in the same cluster, $C_p$ -- centered
at prefix $p$. We call $p$ the \emph{pivot} of cluster $C_p$. Then
the prefixes that are assigned to $C_p$ are removed from the set of prefixes $P$ 
and the {\pivot} algorithm is reapplied to the remaining subset of prefixes 
that have not been assigned to any cluster.

\begin{figure}[tbp]
\centerline{
\includegraphics[width=0.42\textwidth]{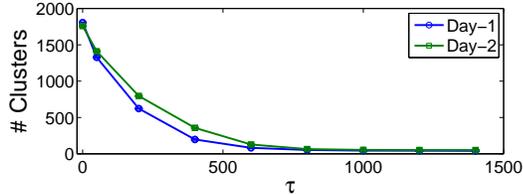}}
\caption{Number of clusters vs.\ $\tau$} 
\vspace{-1.2em}
\label{fig:numClusters}
\end{figure}

\mpara{Scaling with \RSclustering.} 
Notice that \pivot\ algorithm is a nonparametric algorithm where the number of 
clusters are not pre-set. In fact, the number of clusters is an output which is controlled
by the choice of $\tau$. Setting $\tau$ to a larger value is likely to decrease 
the number of clusters but increase in-cluster \RSD. That property of 
the \pivot\ algorithm provides flexibility in adjusting the scale of the
problem in practice. Figure~\ref{fig:numClusters} shows the number of clusters
we get by applying the algorithm on the 5491 prefixes (\dsjuly) and 3272 prefixes (\dsjan) in our datasets.  
For each choice of $\tau$, we run the algorithm 10 times and show the
average number of clusters for these runs with 95\% confidence interval.
The figure suggests that up to 600, as $\tau$ increases, the number of clusters decreases.
Figure~\ref{fig:rsd_vs_dist} and Figure~\ref{fig:numClusters}, together, show the trade-off
between the number of clusters and the closeness between prefixes within a given cluster.
For the rest of the evaluation in this paper we set $\tau$ to 200. 
This results in around 600 clusters, i.e.\ reduces the scale of the problem by 90\% for \dsjuly\ and
by 82\% for \dsjan.
The impact of $\tau$ and other properties of 
\pivot\ are discussed in \cite{Gursun12, Gionis07} in further detail. 

Next, we analyze the goodness of the clusters. In order to do that, 
we first introduce the following notations. For a client $X_1 \in p_i$,
if $p_i \in C_l$, then we write $X_1 \in C_l$. In addition, we use the
definitions \psdist$_{max}$, \psdist$_{min}$, \psdist$_{avg}$, and their 
\gdist\ counterparts that are introduced in Section~\ref{subsec:rs}.

\mpara{Evaluating \RSclustering\ .}
We evaluate the goodness of a cluster with two metrics: 
(1) \emph{Growth of the diameter:} Diameter of a cluster is the maximum distance between any two clients of the cluster.
We measure the growth of the diameter as the ratio of the maximum server ranking distance between any clients in the RS-cluster over the
maximum diameter of the prefixes that are in that RS-cluster. Diameter growth of a cluster $C_l$, \dg$(C_l)$ is defined below for
\psdist. Note that the \gdist\ counterpart is defined likewise.

\begin{equation}
     \dg_{\psdist}(C_l) = \frac{\underset{X_1, X_2 \in C_l}{\mbox{max}} \; \psdist (x_1,x_2)}{\underset{p_i \in C_l}{\mbox{max}} \; \psdist_{max}(p_i)}
\end{equation}

(2) \emph{Growth of the average pairwise distances:} It is the ratio between average pairwise distances 
within a cluster and the maximum average pairwise distances of its member prefixes. Let $s_l$
be the number of clients in $C_l$. Average pairwise growth of a cluster $C_l$, \ag$(C_l)$, is defined below for \psdist. 
Note that the \gdist\ counterpart is defined likewise.

\begin{equation}
     \ag_{\psdist}(C_l) = \frac{ \frac{1}{s_l * (s_l-1)} \sum_{X_1, X_2 \in C_l} \; \psdist (x_1,x_2)} {\underset{p_i \in C_l}{\mbox{max}} \; \psdist_{avg}(p_i)}
\end{equation}

\begin{figure*}[tbp]
\centerline{
\includegraphics[width=0.40\textwidth]{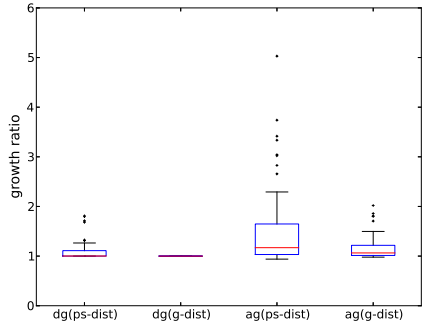}
\includegraphics[width=0.40\textwidth]{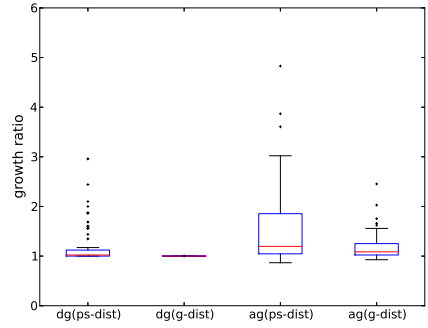}}
  \centerline{\hfill\hfill(a) \dsjuly\ \hfill\hfill\hfill(b) \dsjan\ \hfill\hfill}
\caption{In each box, the red line is the median distance, 
the upper and the lower end of the box represent the first and third quantile, respectively.
The upper and lower ends of the whiskers represent the maximum and minimum values in 
the data, respectively. The individual red points are outliers.} 
\label{fig:inClustEval}
\end{figure*}

By definition, when \dg\ and \ag\ values for a cluster $C_l$ are near 1, 
it means the diameter and average pairwise distances within $C_l$ 
did not grow further than the diameter and the average pairwise distances 
of its least compact prefix, respectively.
Figure~\ref{fig:inClustEval} shows \dg\ and \ag\ statistics for each cluster. 
The figure shows that the \dg\ and \ag\ values for both \psdist\ and \gdist\ are around 1,
that is the clusters did not grow out of their prefixes significantly. That means \RSclustering\
generates compact clusters with respect to its data units.

\begin{figure*}[tbp]
\centerline{
\includegraphics[width=0.55\textwidth]{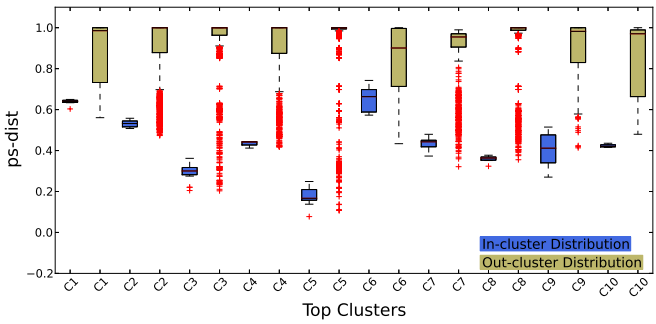}
\includegraphics[width=0.55\textwidth]{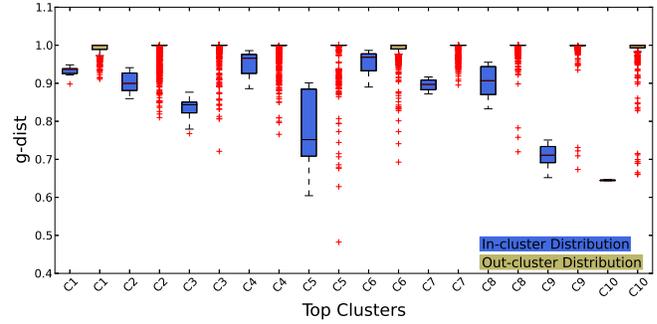}}
\vspace{-1.2em}
\caption{(a) \psdist\ (b) \gdist. In each box, the red line is the median distance, 
the upper and the lower end of the box represent the first and third quantile, respectively.
The upper and lower ends of the whiskers represent the maximum and minimum values in 
the data, respectively. The individual red points are outliers.} 
\label{fig:largeRSDCluster_inOut_dist}
\end{figure*}

\begin{figure*}[tbp]
\centerline{
\includegraphics[width=0.40\textwidth]{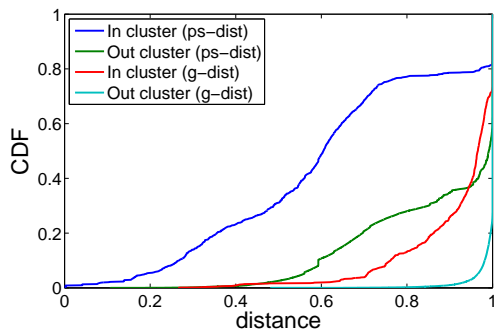}
\includegraphics[width=0.40\textwidth]{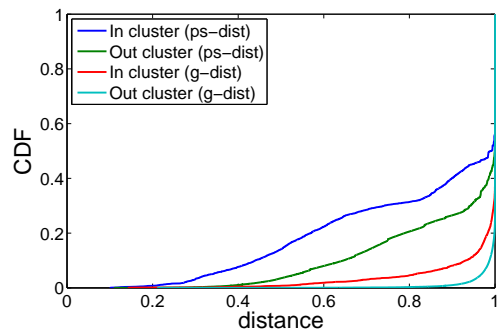}}
  \centerline{\hfill\hfill(a) \dsjuly\ \hfill\hfill\hfill(b) \dsjan\ \hfill\hfill}
\caption{in-cluster vs. out-cluster distances} 
\vspace{-1em}
\label{fig:avg_inOut}
\end{figure*}

Next, in order to test the goodness of clusters further, we compare the average pairwise 
\psdist\ and \gdist\ across prefixes that belong to the same cluster with the ones that belong to 
different clusters. We call the former in-cluster and the latter out-cluster distances. 
Figure~\ref{fig:avg_inOut} shows that there is a great difference in the 
distribution of distances between prefixes that are from the same cluster compared to the
ones that are from different clusters.

In order to investigate the cluster statistics even further, we look at top 10
clusters in more detail. For each prefix $p_1 \in C_l $, we compute 
$\psdist_{avg}(p_1, p_2)$ for all $p_2 \in C_l$. Also, we compute   
$\psdist_{avg}(p_1, p_3)$ for all $p_3 \notin C_l$. We call the 
the former in-cluster, and the latter out-cluster. The hypothesis 
is that the average distance of a prefix should be much closer to 
another prefix that is in the same cluster compared to the ones
that are outside the cluster. We compute in-cluster and 
out-cluster averages for all prefixes in the largest 10 clusters from \dsjuly\ and plot the
statistics in Figure~\ref{fig:largeRSDCluster_inOut_dist}. The figure shows
that across all clusters, in-cluster distances are lower compared to the out-cluster
distances.

In addition Table~\ref{tab:inOut} shows some statistics for these 10 clusters including 
their sizes (the number of prefixes per cluster), average \RSD\ values between 
prefixes, the number of unique ASes that the prefixes in the clusters belong to, 
and the geo locations of the prefixes in each cluster.
One thing to note is that for 7 out of these 10 clusters, the prefixes are from one single country. 
Also note that, the variety of ASes within a cluster ranges from 1 to 23. 
In fact, looking closer into the set of all clusters, we find that  80\% of the clusters 
are composed of prefixes from the same country and 38\% of them are 
composed of prefixes from the same AS. 
To that end we ask the following question next : how does \RSclustering\ compares
with clustering prefixes by their ASes or geographic locations ?

 \begin{table*}
   \centering
   \topcaption{Statistics for 10 large clusters of \dsjuly.}
\scalebox{0.90}{
   \begin{tabular}{@{}l|rrrrrrrrrrrrrrrrrrrrrrrrrrrrrr@{}}
 & \multicolumn{3}{l}{\small $C_1$} & \multicolumn{3}{l}{\small $C_2$} &
 \multicolumn{3}{l}{\small $C_3$} & \multicolumn{3}{l}{\small $C_4$} &
 \multicolumn{3}{l}{\small $C_5$} & \multicolumn{3}{l}{\small $C_6$} &
 \multicolumn{3}{l}{\small $C_7$} & \multicolumn{3}{l}{\small $C_8$} &
 \multicolumn{3}{l}{\small $C_9$} & \multicolumn{3}{l}{\small $C_{10}$}  \\
 \midrule
\small Size of cluster ($C$) & \multicolumn{3}{l}{\small 102} &
\multicolumn{3}{l}{\small 102} & \multicolumn{3}{l}{\small 102} &
\multicolumn{3}{l}{\small 85} & \multicolumn{3}{l}{\small 64} &
\multicolumn{3}{l}{\small 52} & \multicolumn{3}{l}{\small 48} &
\multicolumn{3}{l}{\small 35} & \multicolumn{3}{l}{\small 26} &
 \multicolumn{3}{l}{\small 18}  \\
\small Avg. \RSD\ ($C$) & \multicolumn{3}{l}{\small 59.73} &
\multicolumn{3}{l}{\small 398.36} & \multicolumn{3}{l}{\small 16.99} &
\multicolumn{3}{l}{\small 167.31} & \multicolumn{3}{l}{\small 108.27} &
\multicolumn{3}{l}{\small 83.88} & \multicolumn{3}{l}{\small  93.51} &
\multicolumn{3}{l}{\small 42.65} & \multicolumn{3}{l}{\small 31.42} &
 \multicolumn{3}{l}{\small 88.39}  \\ 
\small Countries & \multicolumn{3}{l}{\small FR} &
 \multicolumn{3}{l}{\small CH} &  \multicolumn{3}{l}{\small DE} &
  \multicolumn{3}{l}{\small DE} & \multicolumn{3}{l}{\small FR,ES,IT,BE} &
  \multicolumn{3}{l}{\small FR} & \multicolumn{3}{l}{\small DE} &
  \multicolumn{3}{l}{\small DE} & \multicolumn{3}{l}{\small DE,IT} &
  \multicolumn{3}{l}{\small FR,DE} \\ 
  \small num. unique ASes & \multicolumn{3}{l}{\small {23}} &
 \multicolumn{3}{l}{\small 6} &  \multicolumn{3}{l}{\small 10} &
  \multicolumn{3}{l}{\small 16} & \multicolumn{3}{l}{\small 17} &
  \multicolumn{3}{l}{\small 1} & \multicolumn{3}{l}{\small 7} &
  \multicolumn{3}{l}{\small 4} & \multicolumn{3}{l}{\small 5} &
  \multicolumn{3}{l}{\small 2} \\ 


\bottomrule
 \end{tabular}
}
 \label{tab:inOut}
 \end{table*}

\subsection{\RSclustering\ vs. clustering by AS and Geography}
\label{subsec:clust-asgeo}

\begin{figure*}[tbp]
\centerline{
\includegraphics[width=0.33\textwidth]{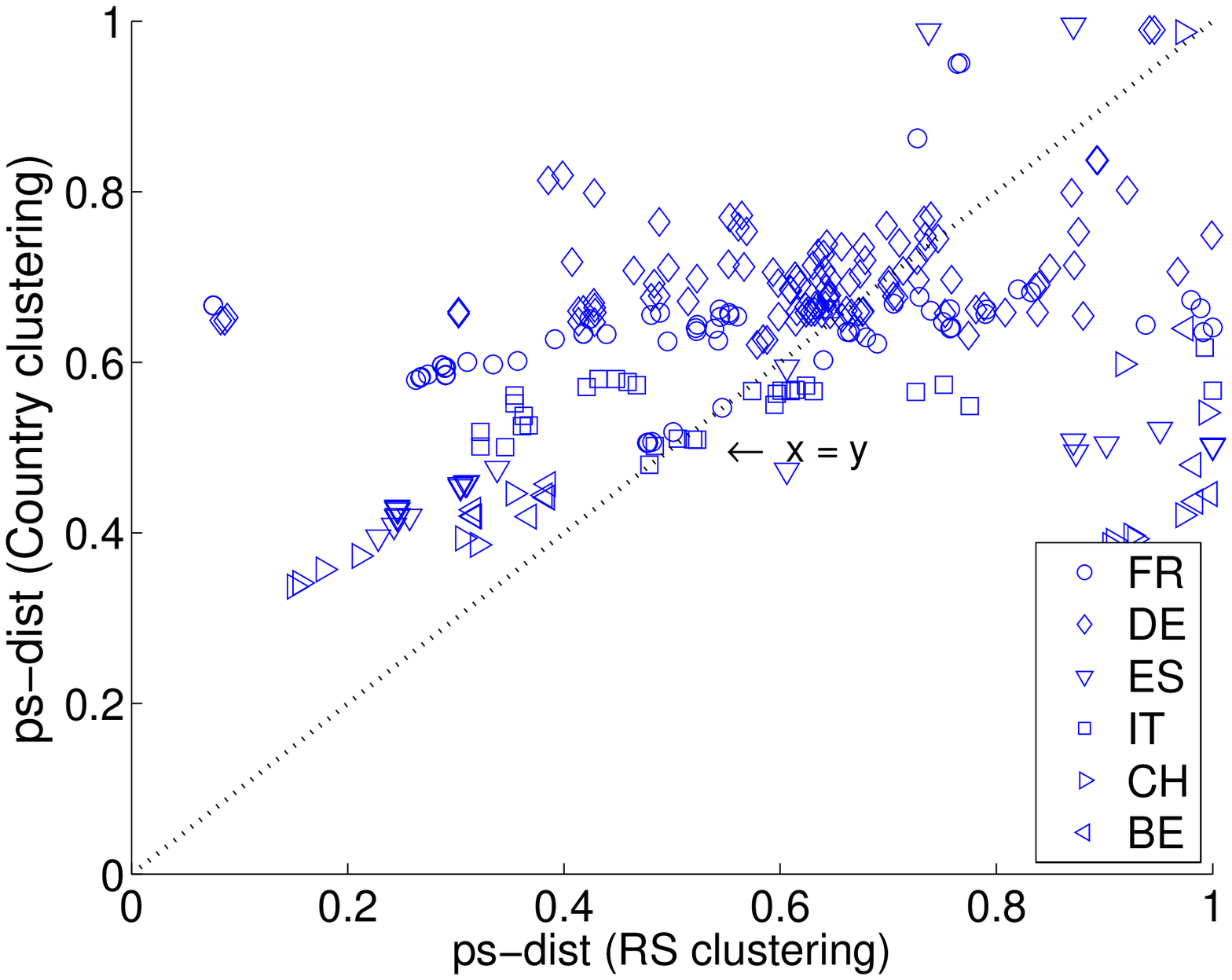}
\includegraphics[width=0.33\textwidth]{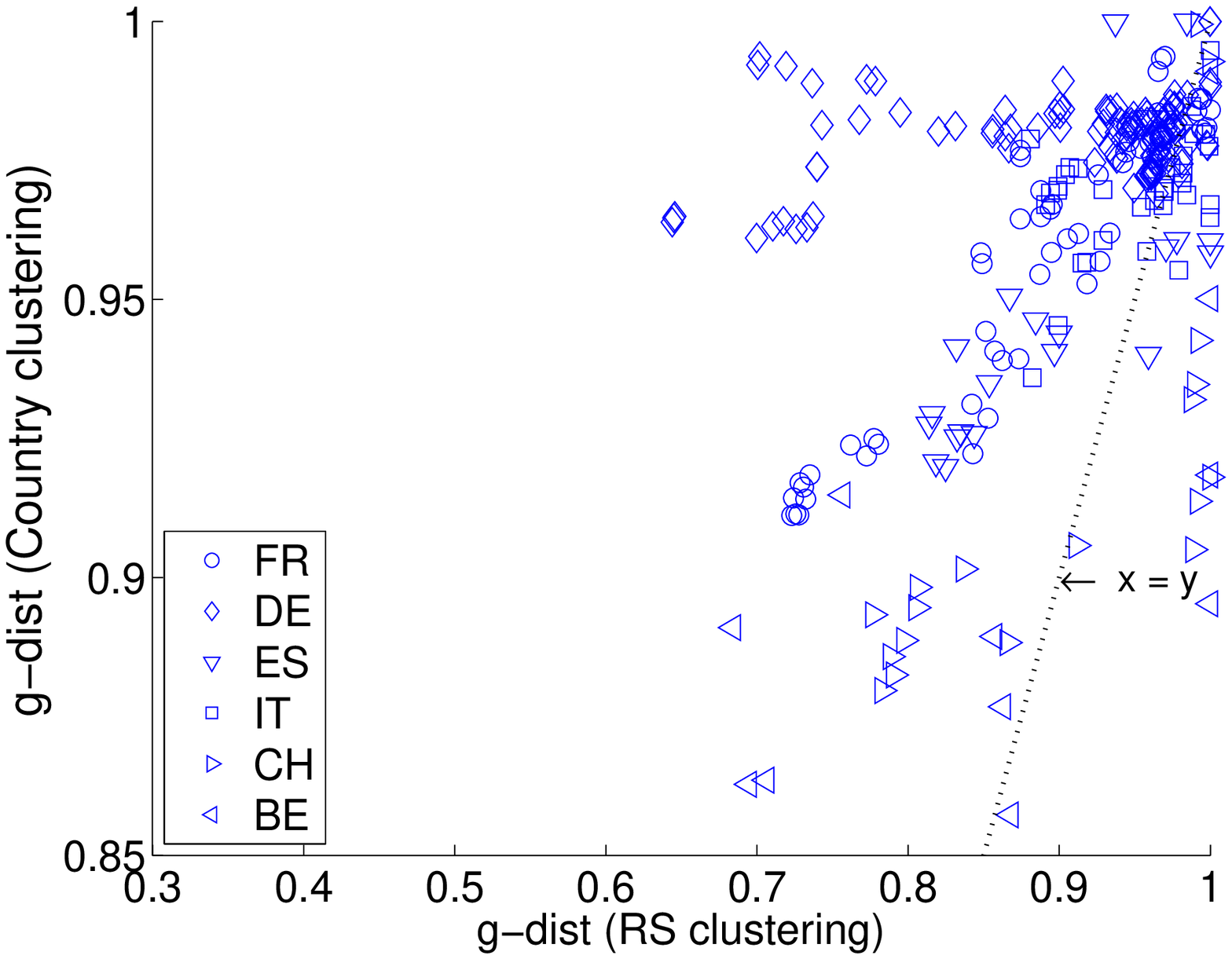}
\includegraphics[width=0.33\textwidth]{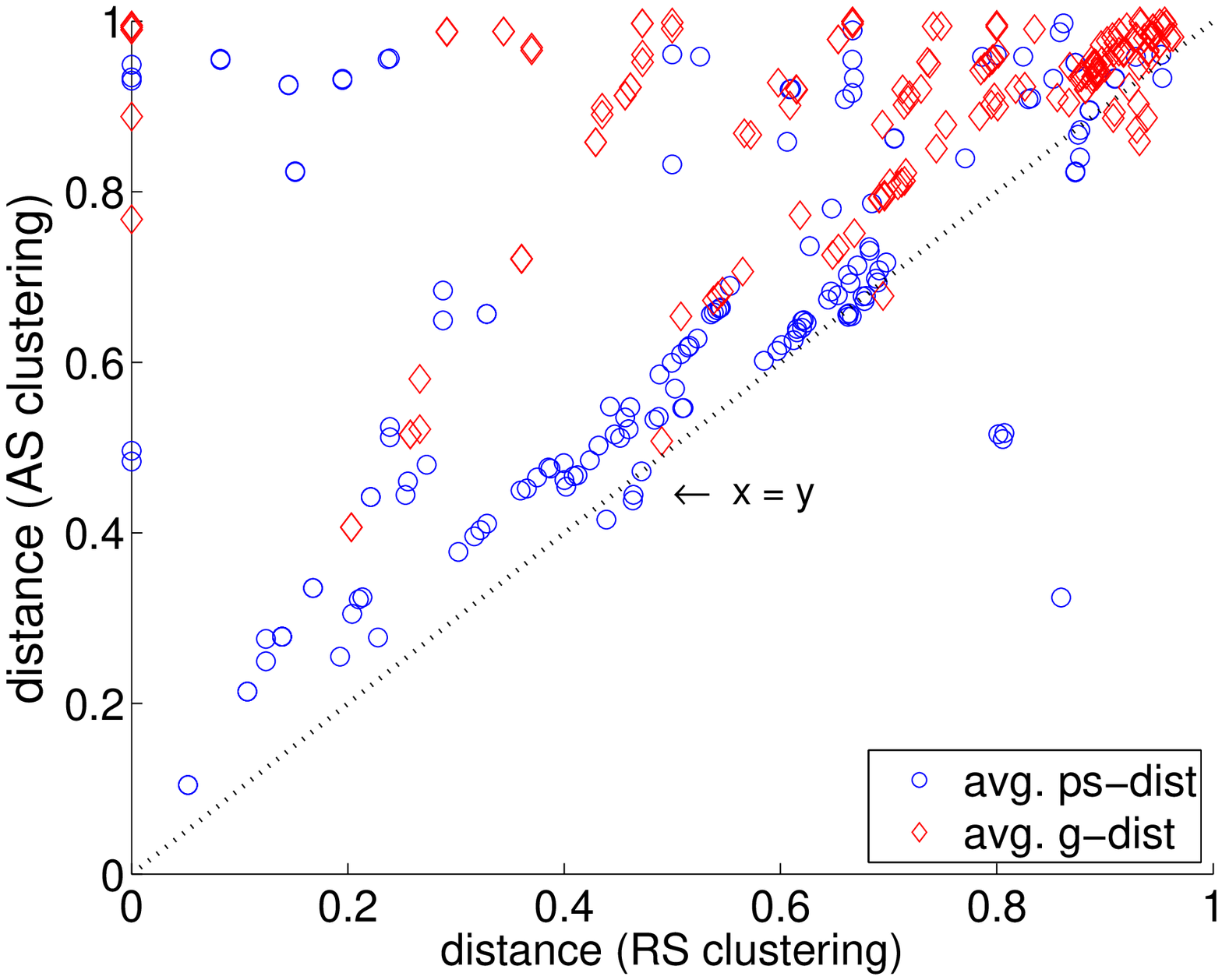}}
  \centerline{\hfill\hfill(a)\hfill\hfill\hfill(b)\hfill\hfill\hfill(c)\hfill\hfill}
\caption{\dsjuly: Comparing \RSclustering\ with clustering by country for (a) \psdist\ (b) \gdist. (c) Comparing \RSclustering\ with clustering by AS for \psdist\ and \gdist.} 
\label{fig:cou_vs_rsd}
\end{figure*}

\begin{figure*}[tbp]
\centerline{
\includegraphics[width=0.33\textwidth]{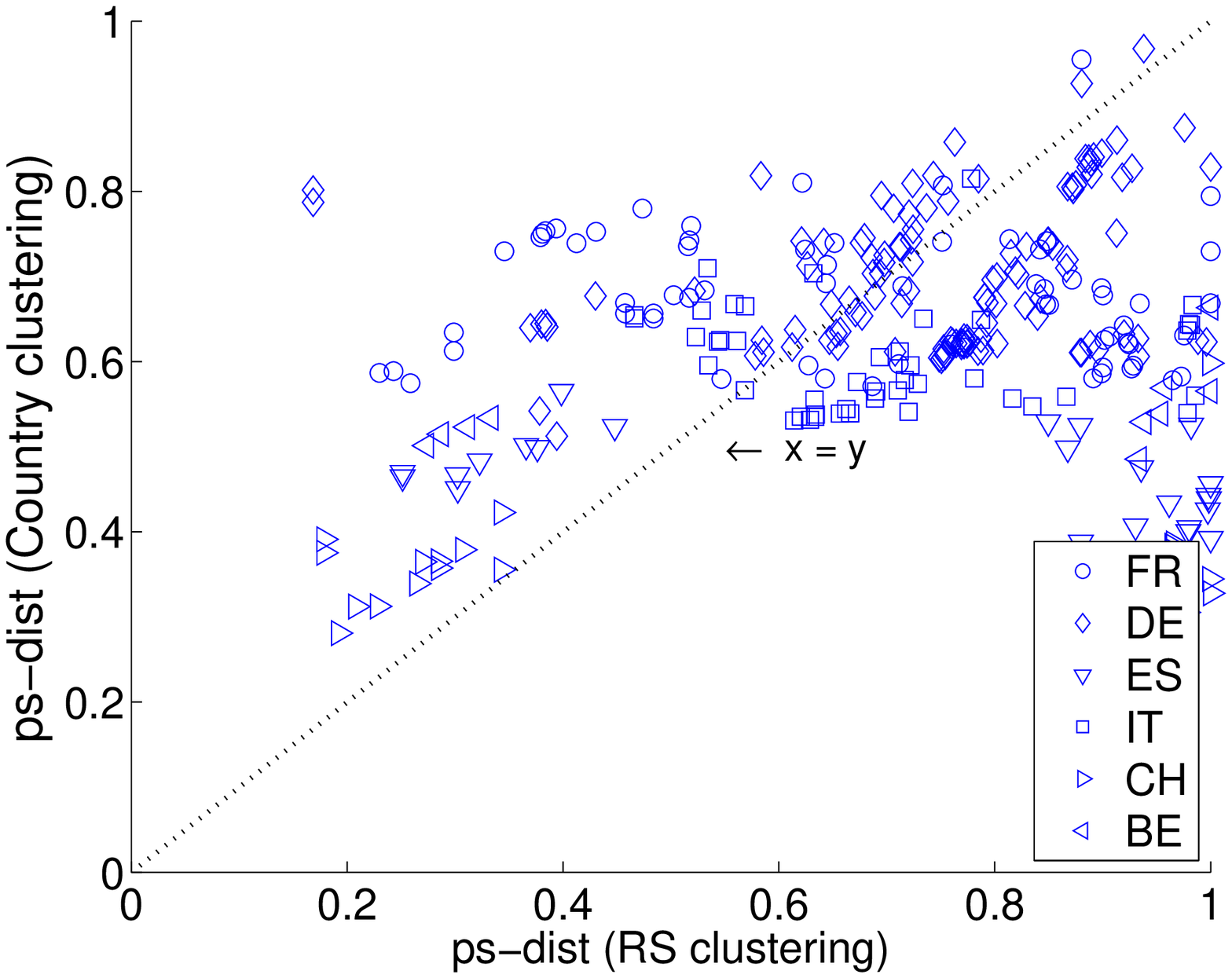}
\includegraphics[width=0.33\textwidth]{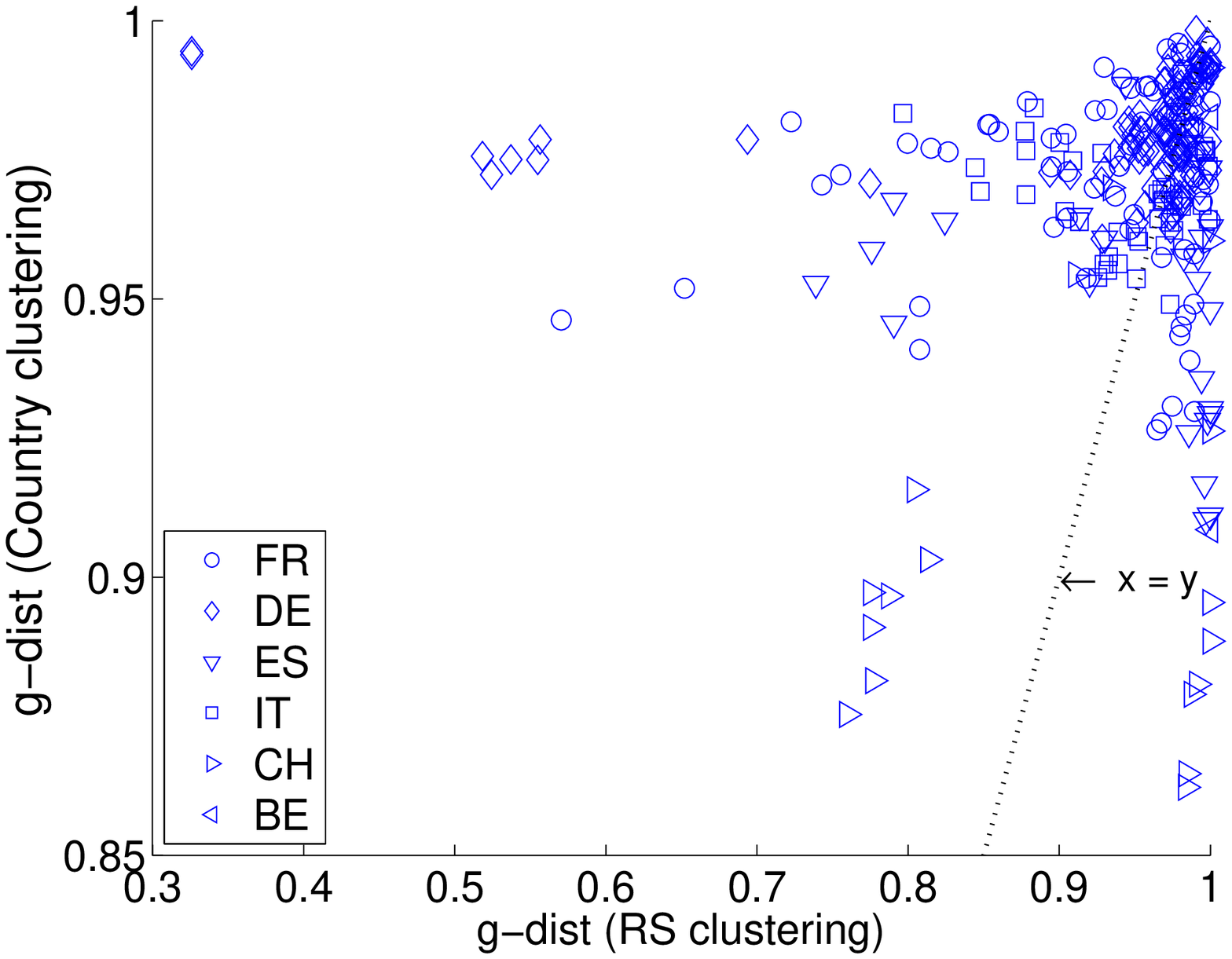}
\includegraphics[width=0.33\textwidth]{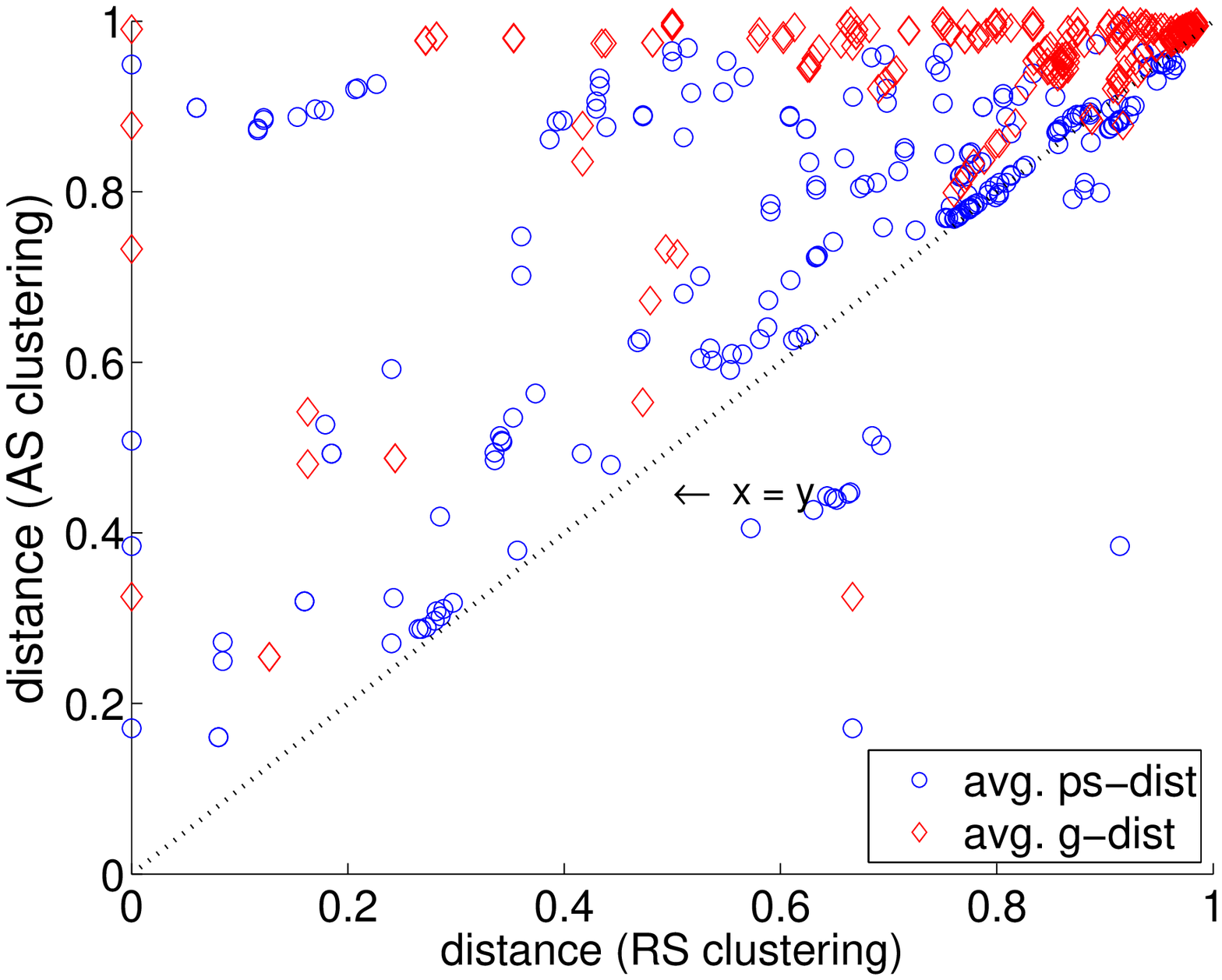}}
  \centerline{\hfill\hfill(a)\hfill\hfill\hfill(b)\hfill\hfill\hfill(c)\hfill\hfill}
\caption{\dsjan: Comparing \RSclustering\ with clustering by country for (a) \psdist\ (b) \gdist. (c) Comparing \RSclustering\ with clustering by AS for \psdist\ and \gdist.} 
\label{fig:cou_vs_rsd2}
\end{figure*}

First we compare \RSclustering\ with clustering by country. We map
each prefix to its country\footnote{We use Akamai's EdgeScape tool for IP to geo mapping \cite{edgescape}} . 
Then, for each prefix, we compute $\psdist_{avg}$ and $\gdist_{avg}$ 
with every other prefix in its RS-cluster
and country cluster separately. Then, we take the mean of these averages 
and plot them in Figure~\ref{fig:cou_vs_rsd} (a-b) for \dsjuly\ and 
in Figure~\ref{fig:cou_vs_rsd2} (a-b) for \dsjan. Each point in the figure represents 
a prefix. We plot $x=y$ line for comparison, i.e.\ if the prefix is above the line 
it indicates that the prefix is closer to the other prefixes in its RS-cluster than 
the ones in its country cluster. Figure~\ref{fig:cou_vs_rsd} (a-b) and Figure~\ref{fig:cou_vs_rsd2} (a-b)
show that clustering by routing similarity is not same as clustering by geo location.
They also show that clustering by routing similarity results in better clusters
as the majority of the prefixes are above the $x=y$ line both in \psdist\ and \gdist.



Second we compare \RSclustering\ with clustering by AS. 
We map each prefix to its AS. The set of 5491 prefixes from \dsjuly\ map to 1397 unique ASes. 
The set of 3272 prefixes from \dsjuly\ map to 1206 unique ASes. 
For each prefix, we compute $\psdist_{avg}$ and $\gdist_{avg}$ with every other prefix in its RS-cluster
and AS-cluster separately. Then, we take the mean of these averages 
and plot them in Figure~\ref{fig:cou_vs_rsd} (c) and  Figure~\ref{fig:cou_vs_rsd2} (c). Each point in the figures represents 
a prefix. We plot $x=y$ line for comparison, i.e.\ if the prefix is above the line that indicates 
the prefix is closer to the 
other prefixes in its RS-cluster than the ones in its AS-cluster. 
In Figure~\ref{fig:cou_vs_rsd} (c) and Figure~\ref{fig:cou_vs_rsd2} (c), we see that
almost all prefixes are above the $x=y$ line for both \psdist\ and \gdist. In fact, 
we see a group of prefixes that are on the y-axis. For any one of these prefixes,
the \psdist\ (or \gdist) between itself and the others in its RS-cluster is 0, whereas
the \psdist\ (or \gdist) between itself and the others in its AS cluster is greater than 0.4.

In summary, Figure~\ref{fig:cou_vs_rsd} and Figure~\ref{fig:cou_vs_rsd2} show that \RSclustering\ is different than clustering by
geo and AS. In fact, \RSclustering\ outperforms clustering clients by their countries or ASes.
In addition, by \RSclustering\ we can control the number of clusters to be generated, 
whereas clustering by AS and geo fix the number of clusters by definition. 
In that sense, \RSclustering\  is a flexible and accurate way of clustering.


\begin{figure*}[tbp]
\centerline{
\includegraphics[width=0.25\textwidth]{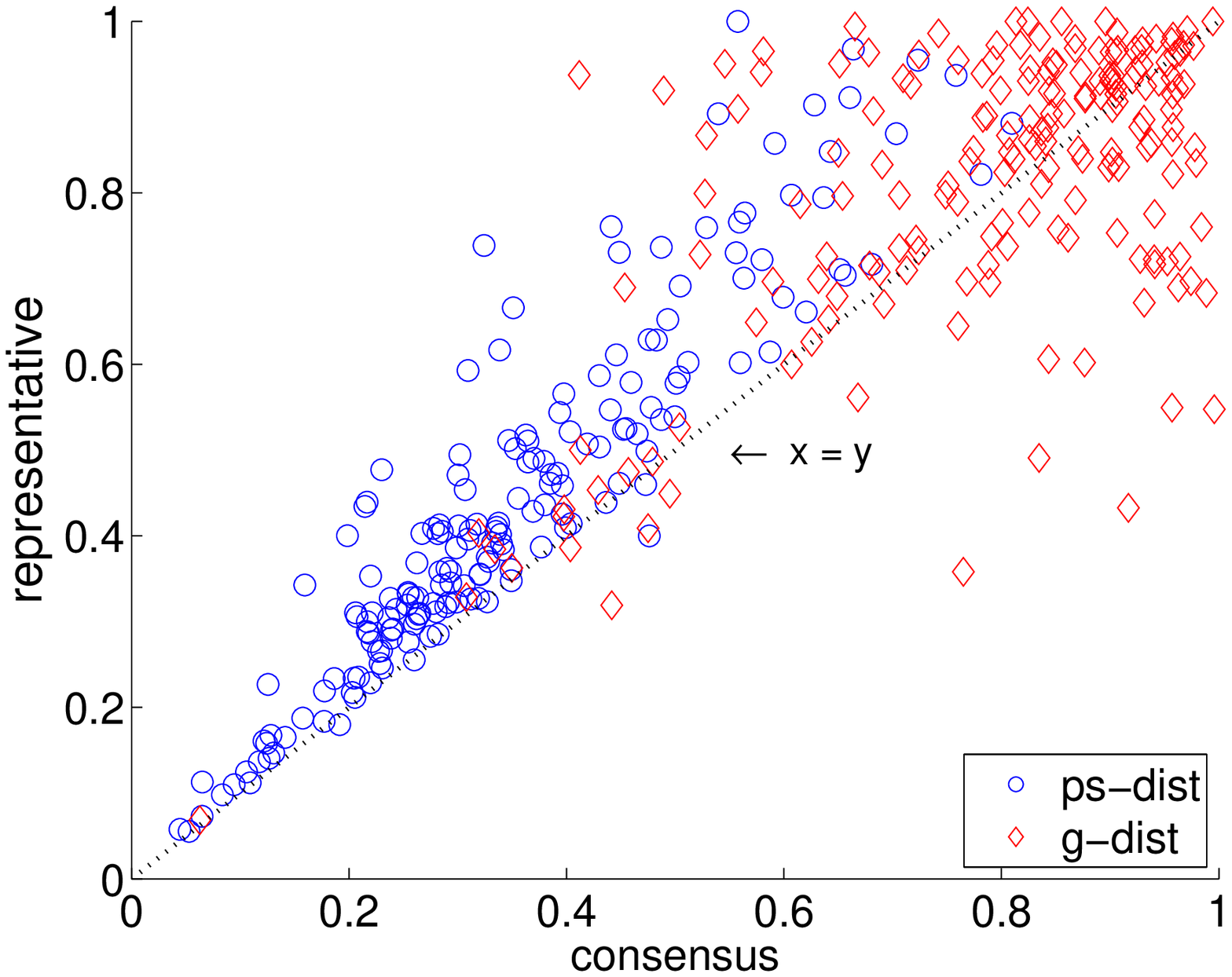}
\includegraphics[width=0.25\textwidth]{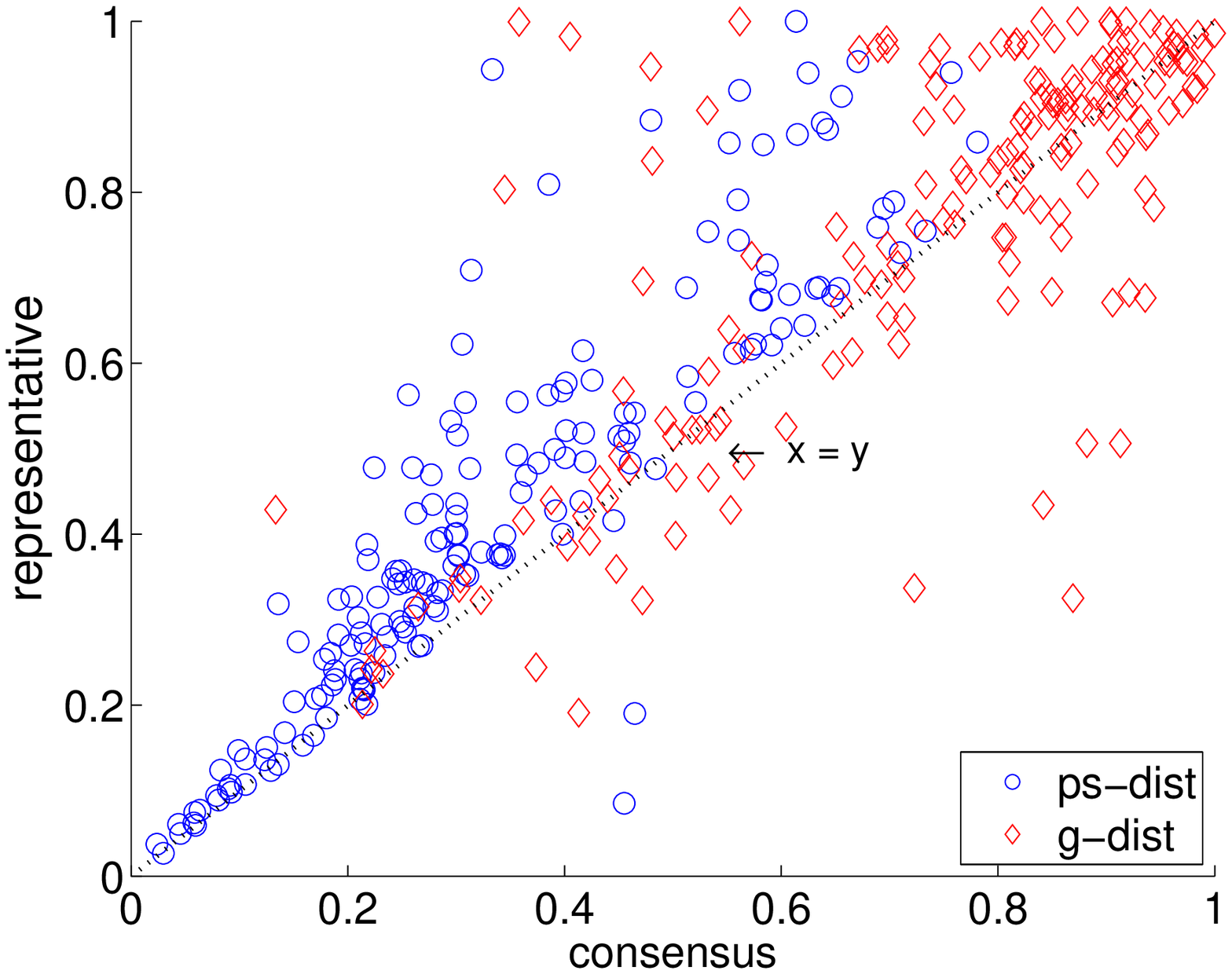}
\includegraphics[width=0.25\textwidth]{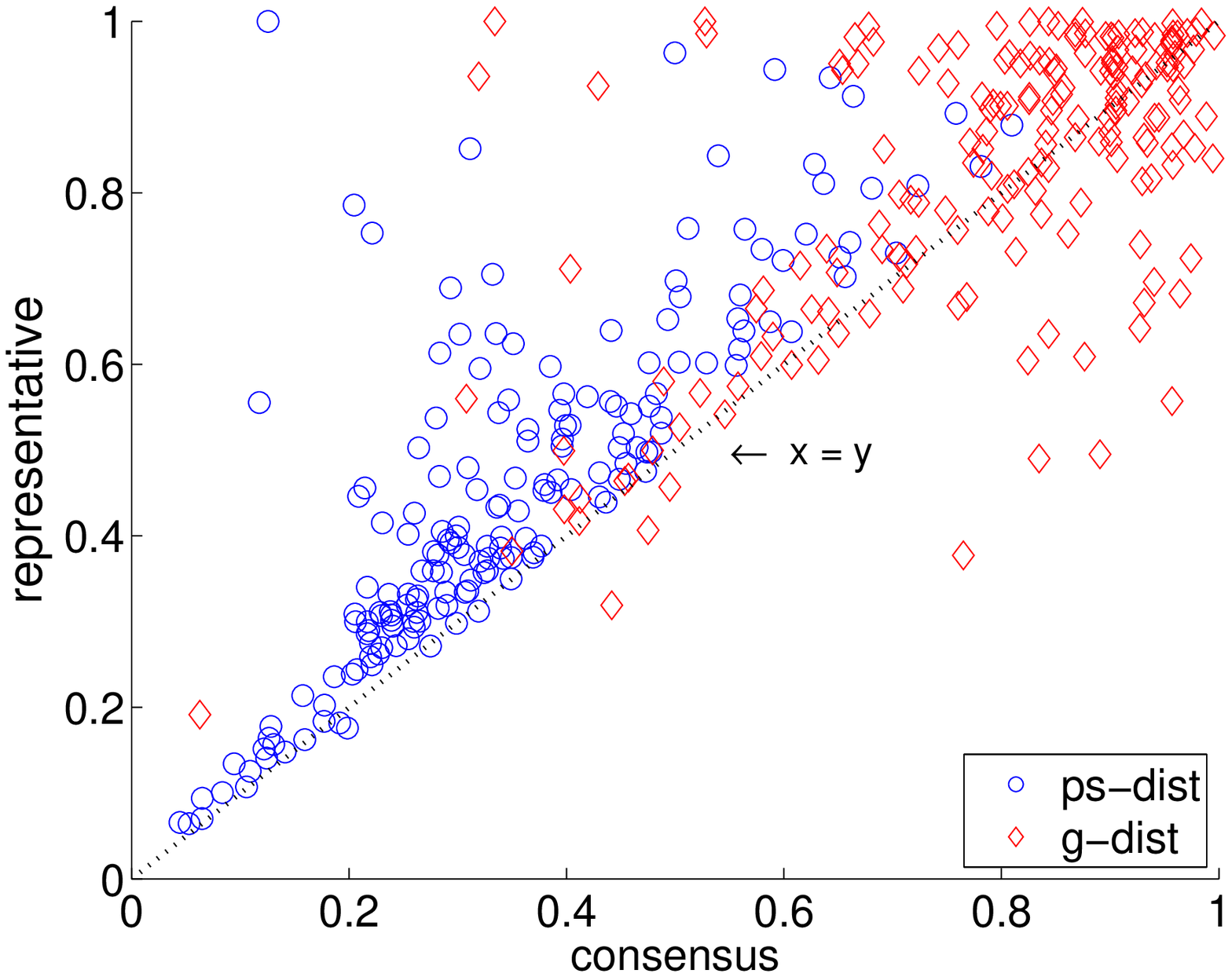}
\includegraphics[width=0.25\textwidth]{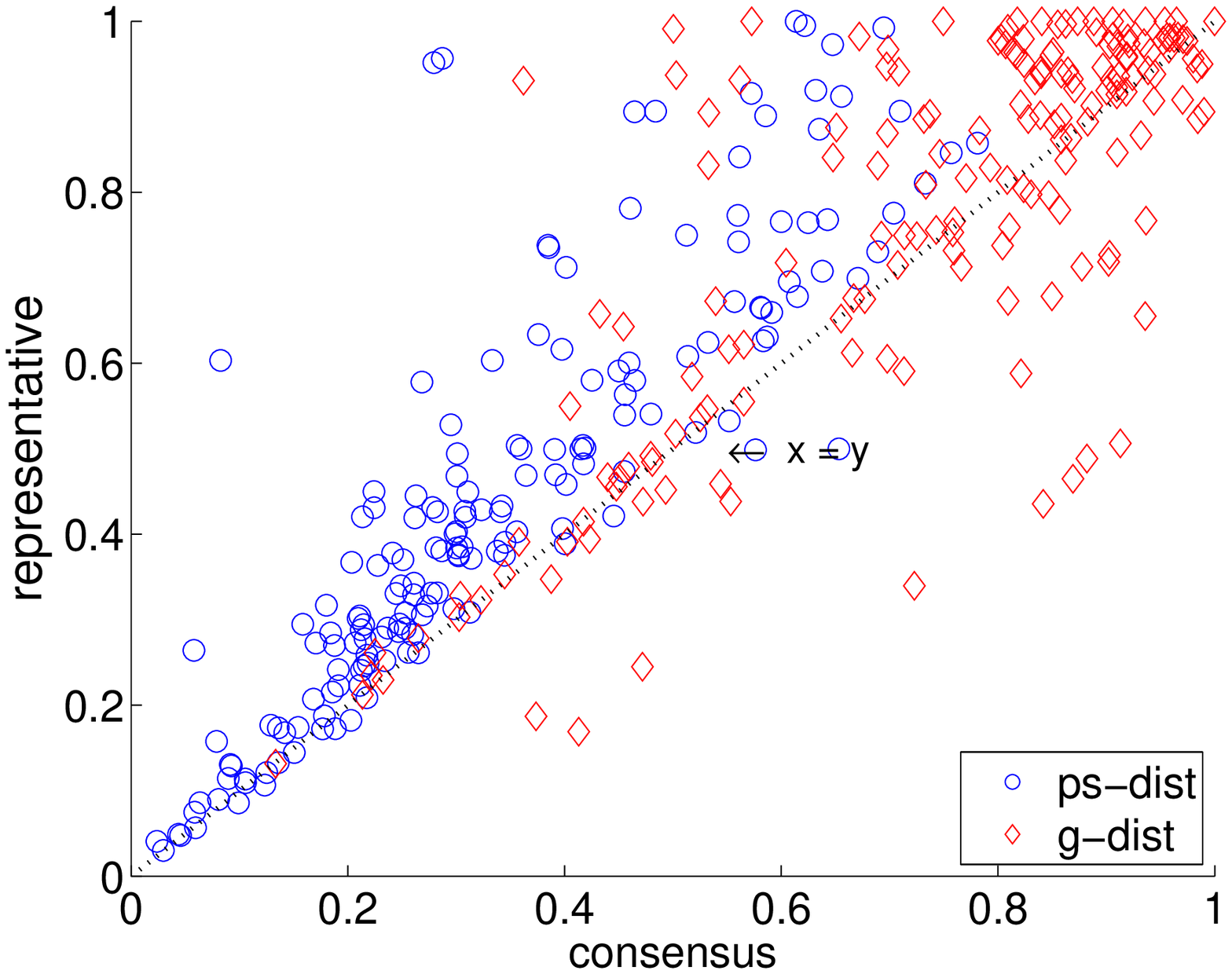}}
  \centerline{\hfill\hfill(a) \dsjuly\hfill\hfill\hfill(b) \dsjan\hfill\hfill\hfill(c) \dsjuly\hfill\hfill\hfill(d) \dsjan\hfill\hfill}
\caption{The comparison of the average distances between (Borda count) consensus vector and all clients in the cluster with the average distances between representative client and all clients in the same cluster.
The representative client is selected at random (a-b) from all clients in the cluster, (c-d) from the pivot prefix.} 
\label{fig:rep_vs_borda}
\end{figure*}

\section{Finding Representative \\ Clients}   \label{sec:application}  Having partitioned the address space, we seek to find a \emph{representative} client from each partition
such that instead of measuring the paths to all clients in the partition, we only measure the ones that are
to the representative. Then we rank the servers according to that representative client. 

One can apply various methods to select a representative. For instance, one method 
is randomly choosing a client from the each cluster. The assumption is that if the cluster is 
compact enough, any client will be a good representative. Second method is choosing a 
client from the \emph{pivot} prefix of each cluster as described in Section~\ref{sec:clust-eval}.
Pivot prefix is the prefix that is the center of its cluster, i.e.\ it is guaranteed that all other
prefixes in the cluster are at most $\tau$ away in \RSD\ space. Given the high 
correlation between \RSD\ and \psdist\ (\gdist) we expect that clients from 
the pivot prefix are good representatives.  
  
In order to investigate the effectiveness of these methods, first we find
a single ranking for each cluster that best describes the rankings of all clients in the cluster.
In other words, we find a \emph{consensus} for each cluster and evaluate the goodness of
the representative client by testing it against the \emph{consensus}.

The problem of aggregating a set of rank vectors and finding consensus is known 
as \emph{rank aggregation} problem \cite{Ailon08, Dwork01}. The problem is
studied extensively and proposed solutions are subject to the definitions of what
properties the consensus should have. For our application, we employ 
two of the proposed solutions, the Borda count \cite{Borda1781} and 
the Plurality method \cite{Dwork01}.

The Borda count is a score-based method. Each candidate's (server's) score is the
sum of the rank values assigned by every voter (client). Once all votes are counted, 
the candidates are reordered from the lowest to the highest rank and the lowest rank candidate
is the winner. The nice property of Borda count is that a Spearman Footrule optimal 
solution can be computed in polynomial time \cite{Dwork01} and therefore it sets a 
fair consensus ranking for our \psdist\ metric.

The Plurality method is another voting scheme where the candidates are 
simply ordered by the number of rank values they are assigned to. For 
instance, a candidate with the most rank 1 assignment is ranked 1, 
similarly, a candidate with the most rank 2 assignment is ranked 2 and so on.
By definition, this method sets a fair consensus for our \gdist\ metric.

To test how good the representatives are, we first compute the consensus 
vector (by both the Borda count and the Plural Method separately). 
Then, we compute the \psdist\ (\gdist) between the 
consensus and all clients in its cluster, and compute their average.
Next, we select a representative for the cluster and we compute the \psdist\ (\gdist) 
between the representative and all clients in its cluster, and compute their average.

Figure~\ref{fig:rep_vs_borda} compares the consensus chosen by the Borda count with
the representatives chosen (a-b) at random from the set of all clients 
in the cluster, and (c-d) at random from the clients that belong to the pivot (center) prefix of each cluster. 
Each point in the figure represents one cluster and the x and y axis are the average distances as 
described above. We also plot $x=y$ line for comparison, i.e.\ if the cluster is close to the line
then it means that the representative client performs as good as the consensus ranking.
The figure shows that most of the clients are very close to the line. The representative client
performs slightly better for \psdist. This is expected by the definition of the Borda count and 
its property of having an spearman foot rule optimal solution.

\begin{figure*}[tbp]
\centerline{
\includegraphics[width=0.25\textwidth]{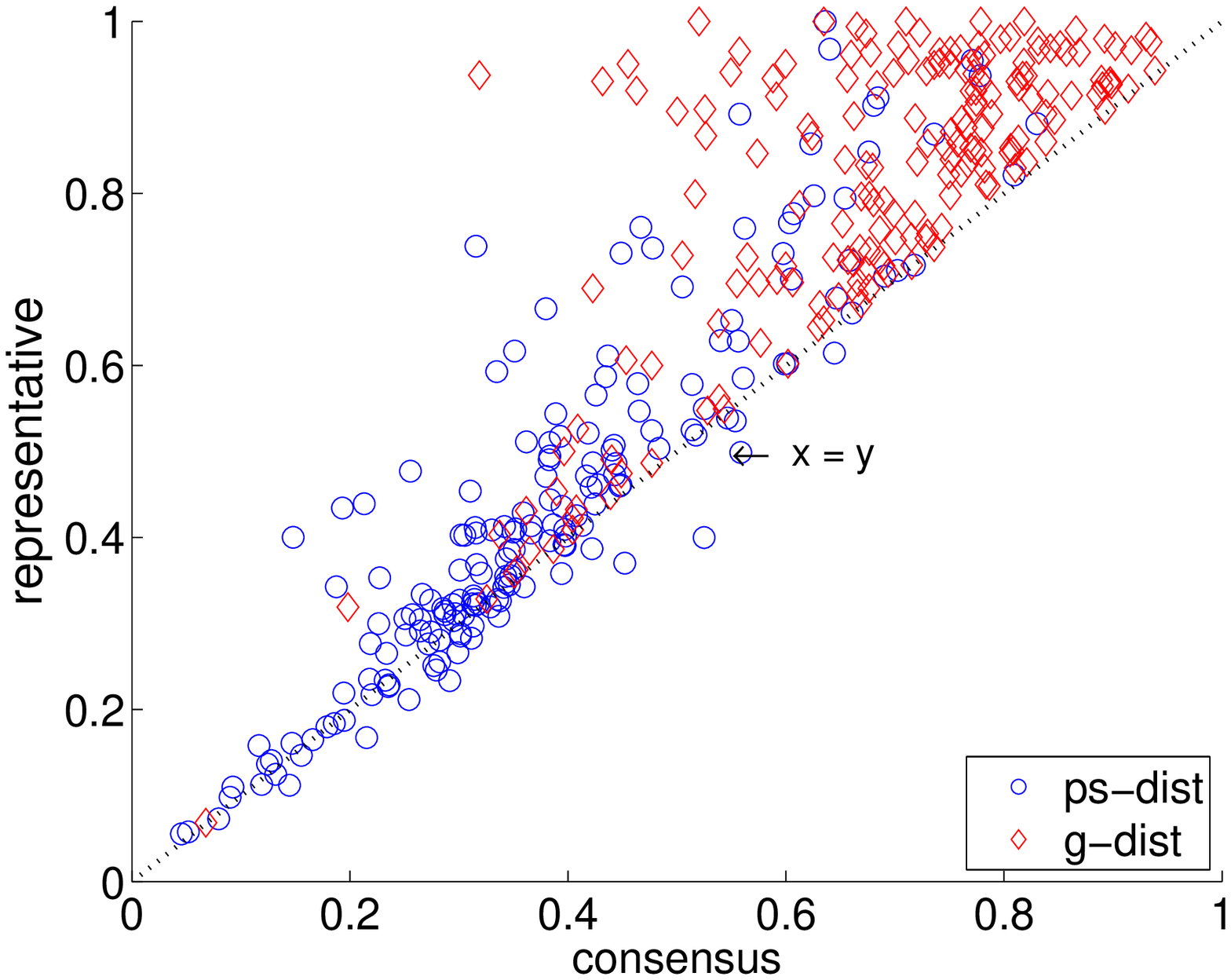}
\includegraphics[width=0.25\textwidth]{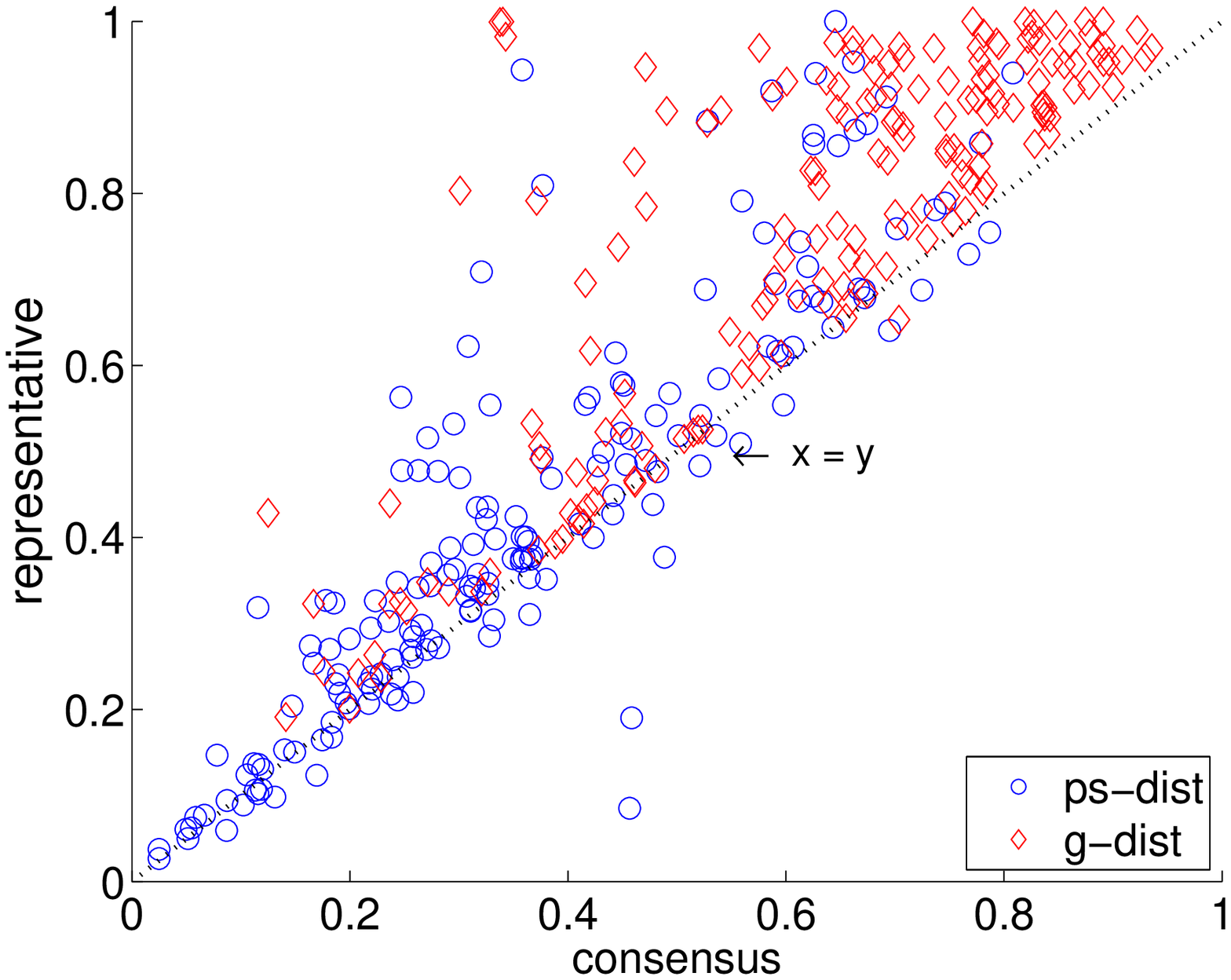}
\includegraphics[width=0.25\textwidth]{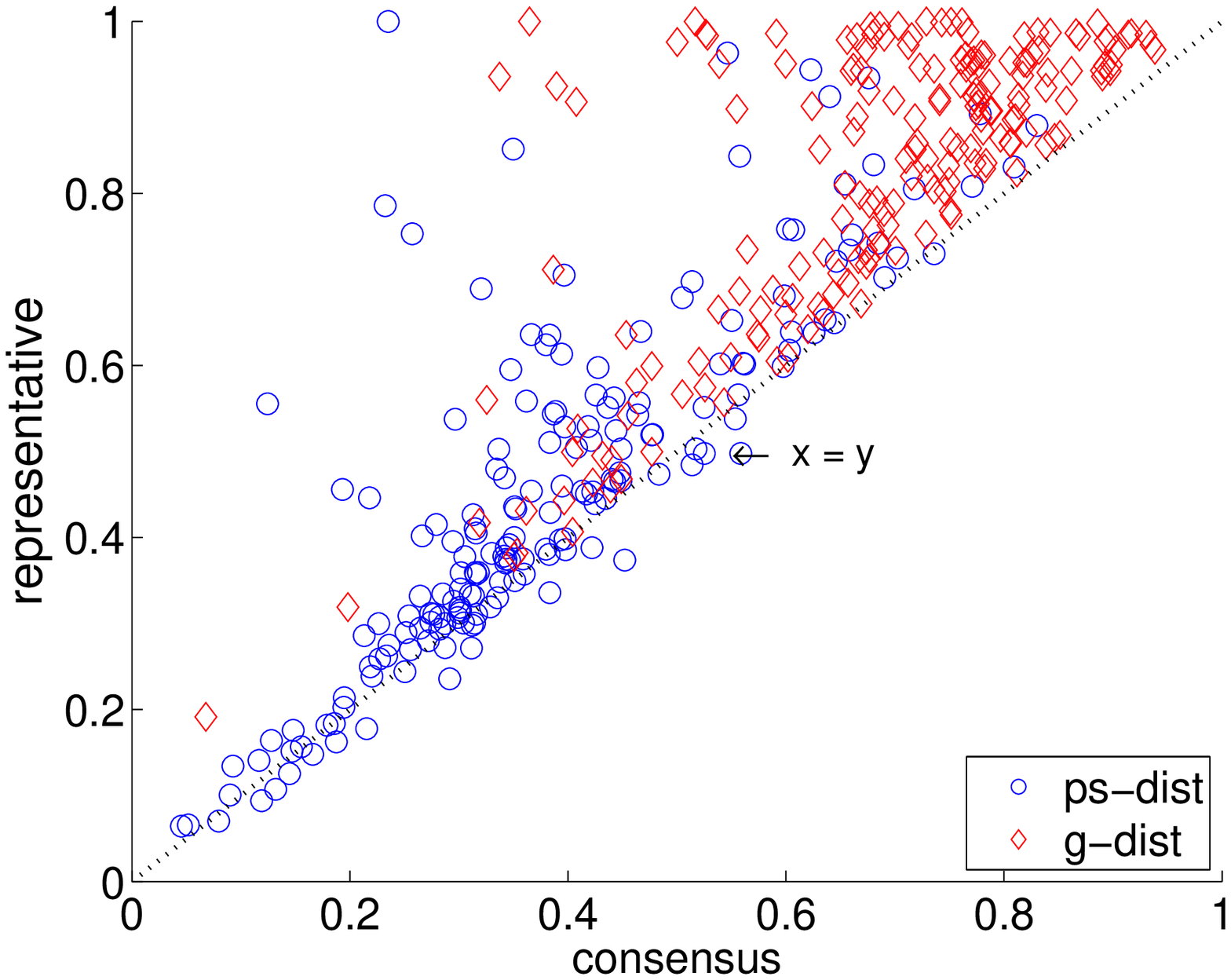}
\includegraphics[width=0.25\textwidth]{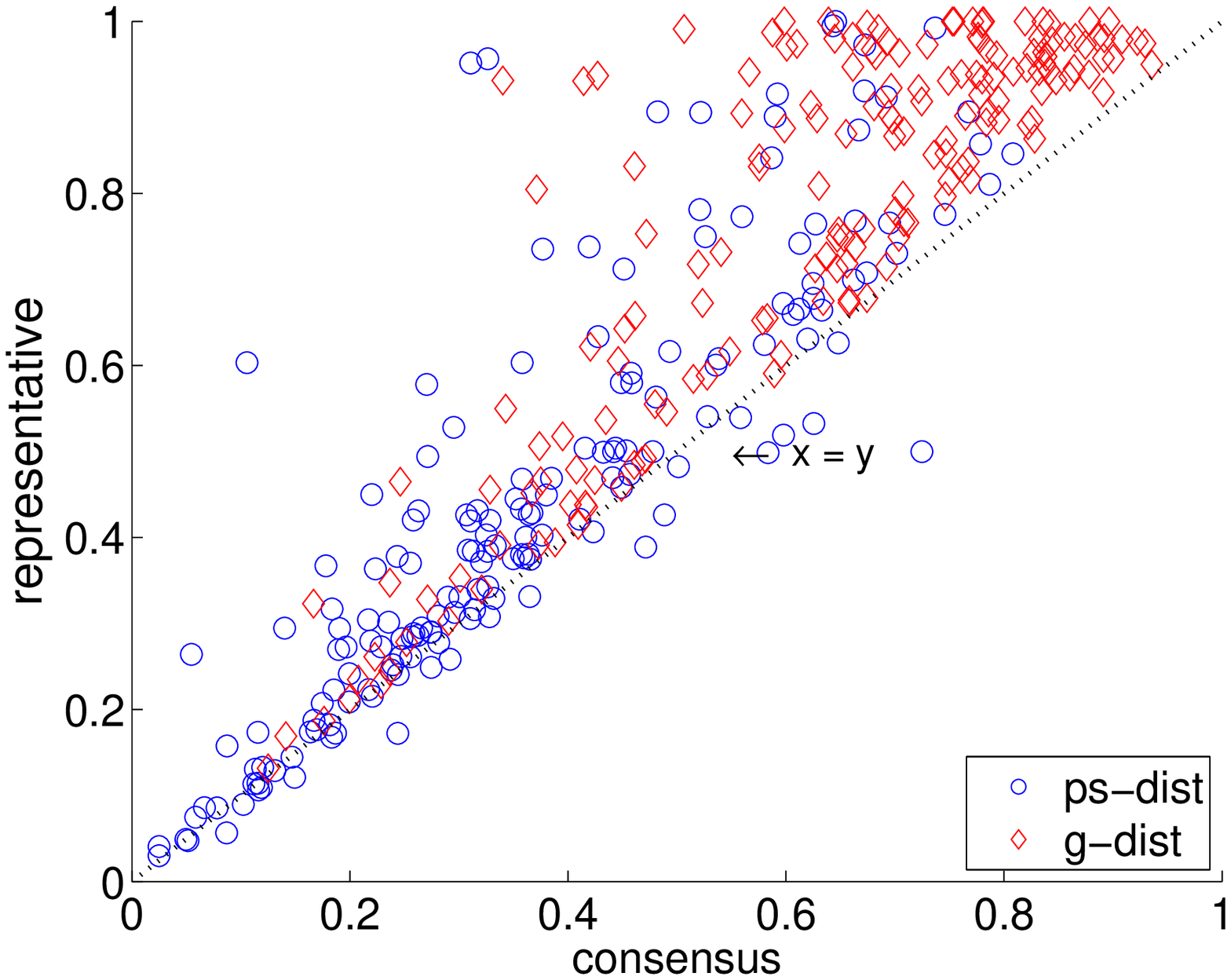}}
  \centerline{\hfill\hfill(a) \dsjuly\hfill\hfill\hfill(b) \dsjan\hfill\hfill\hfill(c) \dsjuly\hfill\hfill\hfill(d) \dsjan\hfill\hfill}
\caption{The comparison of the average distances between (Plural method) consensus vector and all clients in the cluster with the average distances between representative client and all clients in the same cluster.
The representative client is selected at random (a-b) from all clients in the cluster, (c-d) from the pivot prefix.} 
\label{fig:rep_vs_plural}
\end{figure*}

Figure~\ref{fig:rep_vs_plural} is generated exactly the same as described above 
but the consensus is chosen by the Plural method. The figure shows that again, 
most of the clients are very close to the $x=y$ line. The representative clients 
perform better for \gdist\ compared to the Borda count case. This is intuitive by 
the definition of the Plural method which simply orders the servers by the number of 
rank values they are assigned to. In other words, it tends to agree with the
exact rankings of the majority for each position.

\begin{figure*}[tbp]
\centerline{
\includegraphics[width=0.40\textwidth]{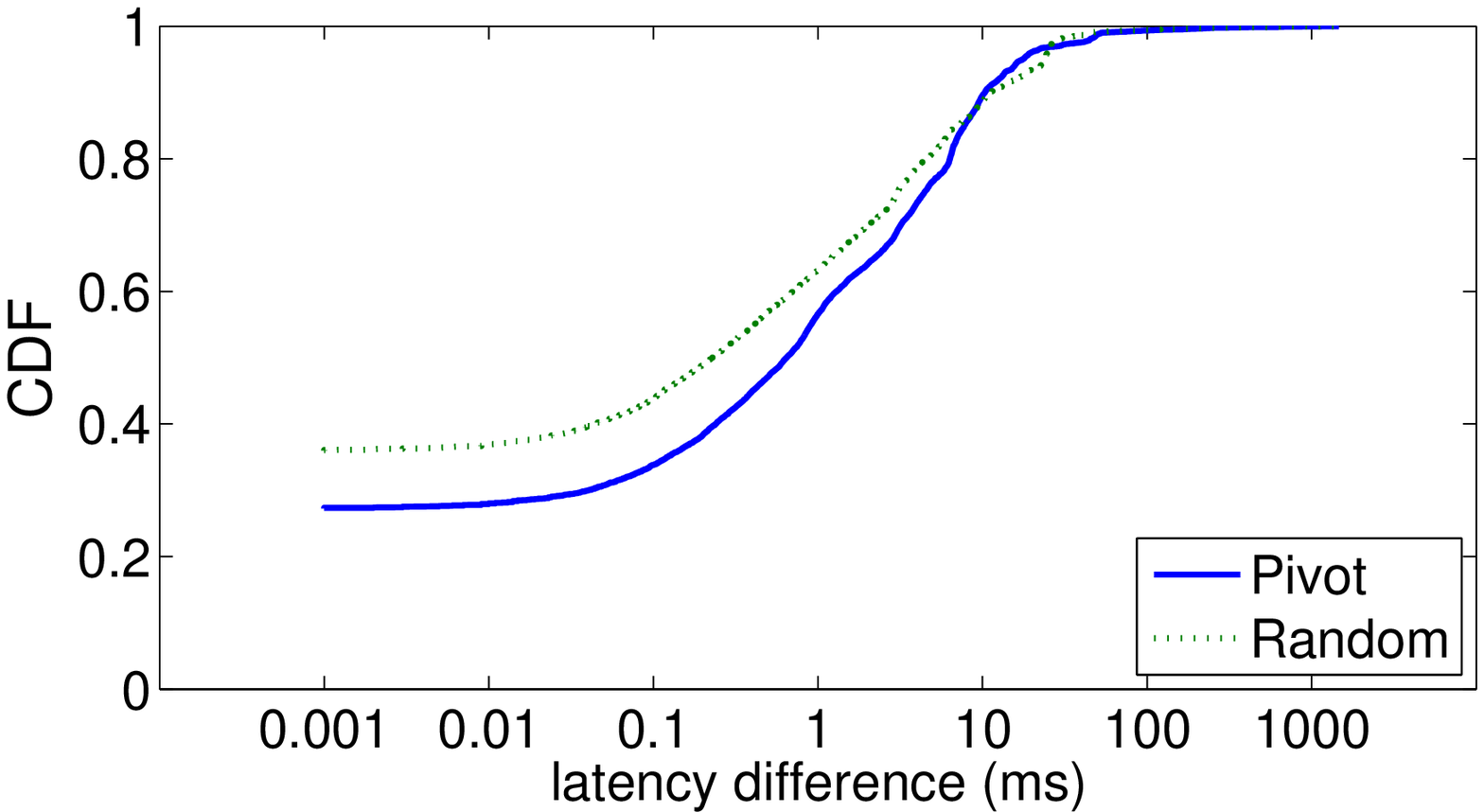}
\includegraphics[width=0.40\textwidth]{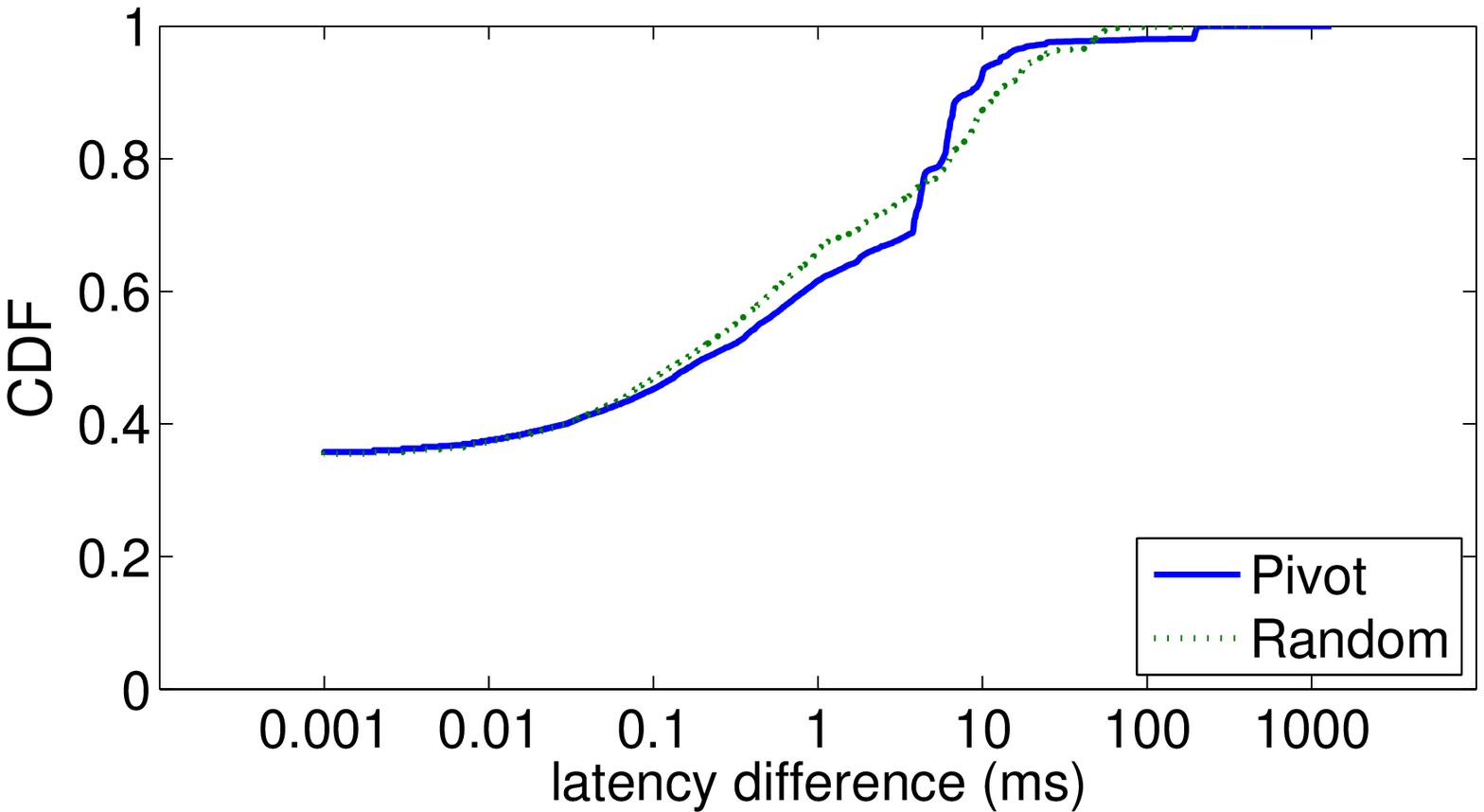}}
  \centerline{\hfill\hfill(a) \dsjuly\ \hfill\hfill\hfill(b) \dsjan\ \hfill\hfill}
\caption{The distribution of latency differences due to assigning top-1 server of the representative } 
\label{fig:app_latDiff}
\end{figure*}

Next we investigate how large the latency can get for a client due to clustering. 
We compute the latency difference for each client in a given cluster as follows. For a given cluster, 
let $T_j$ be the top-1 server with respect to the cluster's representative client. Then, for any other client 
in the cluster, say $X_i$, we compute how much the latency to $X_i$ increases if $T_j$ is assigned to $X_i$ 
instead of its top-1 server.
Figure~\ref{fig:app_latDiff} shows the distribution of such latency for all clients in the clusters.
The figure shows that selecting representatives both random from the clients in the pivot prefix or any client in 
the cluster results in small latency difference for the rest of the clients. In fact, selecting representative
at random from all clients slightly outperforms selecting from the pivot for \dsjuly. This is expected since
the pivot prefix is not necessarily the largest one. 
In conclusion, we show that choosing a client at random from a cluster represents
the other clients in the same cluster successfully and reduce the scale of the ranking task. 


\section{Related Work}  \label{sec:rel-work}   Grouping IP addresses has been of interest to various studies. 
\cite{Krishnamurthy00} and \cite{Beverly08} propose that IPs that are numerically 
close to each other to be grouped together.
\cite{Krishnamurthy00} clusters web client IPs by mapping them into 
their longest matching BGP prefixes so that the IPs within a given cluster is 
under the same administrative domain. 
\cite{Beverly08} assumes that IPs that are numerically close to each other 
show similar features (e.g. latency). Based on this assumption this work partitions a subnet as
a supervised learning task so that in each subpartition the feature at interest is minimized. 
This work does not consider BGP routing information which already provides a natural grouping.
\cite{Chen15} proposes grouping end-user IPs by their /24 prefixes in the case of end-user mapping. 
Unlike our work, in \cite{Krishnamurthy00, Beverly08, Chen15}, 
the level of aggregation is limited to the prefix level or below and the state of the inter domain 
routing in the Internet is not considered.

Clustering IP addresses by their geographic locations is another way to partition the address space.
However, \cite{Gueye07}  and  \cite{Poese11} show that identifying the geolocation of an IP block 
accurately is hard. They claim that most geolocation tools are only reliable at the country level.
Moreover, even in the case where geolocation based grouping is possible, it does not generate the
desired clustering results we discuss in this paper, i.e.\ geographic closeness 
does neither infer topological closeness \cite{Freedman05} nor low latency between servers 
and clients \cite{Krishnan09}.

\cite{Gursun12} proposes the \RSD\ metric and the \pivot\ clustering to group prefixes according 
to the BGP path information. This work focuses more on the basic properties of the metric 
and shows how to uncover the factors that drive ASes to choose their interodomain next hops.  
However, unlike our work, this work does not study the goodness of clusters in terms of 
the path performances experienced by the clients in each cluster or
how similarly the clients in a given cluster orders some set of servers.
 
In addition, the correlation between the routing dynamics and end-to-end path performance is widely studied.
The findings in these studies suggest that we use a routing-aware clustering. 
\cite{Savage99} shows the routing and latency relation by comparing the default routing
paths with alternate paths. \cite{Labovitz00} studies that the stability of paths between 
ISPs effect path performance.
\cite{Wang06} shows that packet loss is significantly increased by the routing changes. 
\cite{Paxson96} and \cite{Feamster03} show that routing instabilities (e.g.\ routing loops and failures) 
can disrupt end to end connectivity.

Incorporating routing information into server redirection problem in CDNs is
studied by \cite{Poese12, Poese12-2}. These works suggest that collaboration between CDNs and ISPs will 
be beneficial for mapping end-users to higher performing CDN servers. To that end, similar to our work, they also 
utilise what can be inferred from routing choices. However, notice that, the problem of server redirection 
studied in  \cite{Poese12, Poese12-2} is different than the server ranking problem that we study in this paper.
Our goal is \emph{not} finding a best server mapping for end-users. Instead, our goal is grouping end-users such that
within each group the order of best to worst performing servers are almost the same. 
In that respect, our work is complementary to \cite{Poese12, Poese12-2}.

\section{Discussion}   \label{sec:discussion}

The \RSclustering\ has properties which makes it desirable for our clustering problem.

First, it is a lightweight algorithm. Notice that both computing pairwise \RSD\ values and running \pivot\ algorithm 
is in $O(n^2)$ where $n$ is the number of prefixes. It is possible to reduce the run time of both 
jobs by pregrouping the prefixes by some coarse grain geographic regions (e.g. their continents or even
countries as we discussed in Section~\ref{subsec:clust-rsd}). For instance, it is very likely that 
for two clients from two different continents, neither their server choices nor the \RSD\ of their prefixes
will be close. Therefore, one can compute \RSD\ and run the \pivot\ separately for 
each coarse geographic area. 

Second, \RSclustering\ is a network-aware method that can capture the dynamicity 
of the Internet, yet it is relatively stable. \cite{Rexford02} shows that 
small fraction of prefixes are responsible for most route changes and 
these are the ones that receive comparatively little traffic, \\
whereas BGP is stable for popular destination prefixes. Moreover, \cite{Comarela13} 
studies the temporal aspects of \RSD\ and show that on any given day, 
approximately 1\% of the next-hop decisions made in the Internet change.
The change goes up to 10\% in a month. This shows that the clusters generated 
by \RSD\ are valid for at least a day (possibly longer). 

Third, \RSclustering\ provides flexibility for adjusting the number of clusters
as we discuss in Section~\ref{subsec:clust-rsd}. The choice of $\tau$ in
\pivot\ effects the number of the clusters generated. In this paper, we set
$\tau$ empirically.
 
Fourth, the only input that \RSclustering\ requires is the set of BGP paths. Notice that
\RSclustering\ does neither rely on the knowledge of the underlying topology nor
the latency on the paths.

Fifth, in our experiments we show that \RSclustering\ reduces the problem
by 90\% for a region in Europe. We note that for other regions in the world, the percentage of
reduction may vary based on how diverse the Internet paths in those regions. Remember that
by definition of \RSD\ the more ASes make similar next hop choices to a set of destination 
prefixes, the smaller their \RSD\ values. The smaller the \RSD\ values are, the more prefixes can be 
grouped together and the more the scale of the problem is reduced. 
Within a given region, one reason that ASes make similar next hop choices is because
the next hop options are limited in the first place. In other words, in regions where path diversity is less, 
we expect to represent prefixes with only a few clusters and therefore reduce the problem significantly.
In our study, we pick a region in Europe where the path diversity is relatively rich compared to the other parts of the world. 
Therefore, we believe that the gain from \RSclustering\ will be more than 90\% in other regions.


One direction for further analysis is the change in \psdist\ and \gdist\ within RS clusters
over time. We believe that such study can have various applications. For instance, one can 
identify the problematic clients/prefixes by observing the unexpected changes in the server
rankings within an RS cluster.

\section{Conclusion}   \label{sec:conc}   In this paper, we introduce a framework to partition the Internet
address space. Our goal is to scale the number of paths to be
monitored between a CDN's servers and its clients. That is 
we aim to find a partitioning of clients such that in each partition
the latencies from servers to the clients in the partition are ordered
similarly. To achieve this goal, we first introduce two metrics (\psdist\ and \gdist)
that measures the similarity between two rank vectors even in the case of 
vectors are known only partially. Second, we show that for any given two clients,
as the number of ASes in the Internet that prefer the same next hops to route to them 
increase, their server ranking similarity increase. Having shown the effect of 
inter domain routing on the server preferences of clients, we employ routing-aware
clustering algorithm. We evaluate the goodness of the clusters by using our metrics
\psdist\ and \gdist\ and show that we obtain compact clusters.
Finally, we show that one can successfully scale the task of server ranking 
by measuring a client at random from each cluster.
\section{Acknowledgements}  \label{sec:ack}  We thank Marcelo Torres, Fangfei Chen, Nick Shectman, Kc Ng, Liang Guo, and Arthur Berger 
from Akamai Technologies for their feedback and insightful discussions. We also thank 
the anonymous referees for their reviews.

\bibliographystyle{plain}
\small
\bibliography{paper}

\begin{thebibliography}{10}

\bibitem{Ailon08}
Nir Ailon, Moses Charikar, and Alantha Newman.
\newblock Aggregating inconsistent information: Ranking and clustering.
\newblock {\em J. ACM}, 55(5):23:1--23:27, November 2008.

\bibitem{Beverly08}
Robert Beverly and Karen Sollins.
\newblock An internet protocol address clustering algorithm.
\newblock In {\em Proceedings of the Third Conference on Tackling Computer
  Systems Problems with Machine Learning Techniques}, SysML'08, pages 5--5,
  Berkeley, CA, USA, 2008. USENIX Association.

\bibitem{Borda1781}
Jean C.~de Borda.
\newblock M{\`e}moire sur les {\`e}lections au scrutin.
\newblock 1781.

\bibitem{Chen15}
Fangfei Chen, Ramesh~K. Sitaraman, and Marcelo Torres.
\newblock End-user mapping: Next generation request routing for content
  delivery.
\newblock In {\em Proceedings of the 2015 ACM Conference on Special Interest
  Group on Data Communication}, SIGCOMM '15, pages 167--181, New York, NY, USA,
  2015. ACM.

\bibitem{Comarela13}
Giovanni Comarela, Gonca G{\"u}rsun, and Mark Crovella.
\newblock Studying interdomain routing over long timescales.
\newblock In {\em Proceedings of the Internet Measurement Conference (IMC)},
  Barcelona, Spain, October 2013.

\bibitem{Diaconis88}
P.~Diaconis.
\newblock {\em {Group representations in probability and statistics.}},
  volume~11 of {\em IMS Lecture Notes-Monograph Series}.
\newblock {Institute of Mathematical Statistics}, 1988.

\bibitem{Dilley02}
John Dilley, Bruce Maggs, Jay Parikh, Harald Prokop, Ramesh Sitaraman, and Bill
  Weihl.
\newblock Globally distributed content delivery.
\newblock {\em IEEE Internet Computing}, 6(5):50--58, September 2002.

\bibitem{google-dns}
Google~Public DNS.
\newblock https://developers.google.com/speed/public-dns/.

\bibitem{open-dns}
Open DNS.
\newblock https://www.opendns.com.

\bibitem{Dwork01}
Cynthia Dwork, Ravi Kumar, Moni Naor, and D.~Sivakumar.
\newblock Rank aggregation methods for the web.
\newblock In {\em Proceedings of the 10th International Conference on World
  Wide Web}, WWW '01, pages 613--622, New York, NY, USA, 2001. ACM.

\bibitem{edgescape}
Akamai EdgeScape.
\newblock http://www.akamai.com/dl/brochures/edgescape.pdf.

\bibitem{Fagin03}
Ronald Fagin, Ravi Kumar, and D.~Sivakumar.
\newblock Comparing top k lists.
\newblock In {\em Proceedings of the Fourteenth Annual ACM-SIAM Symposium on
  Discrete Algorithms}, SODA '03, pages 28--36, Philadelphia, PA, USA, 2003.
  Society for Industrial and Applied Mathematics.

\bibitem{Feamster03}
Nick Feamster, David~G Andersen, Hari Balakrishnan, and M~Frans Kaashoek.
\newblock Measuring the effects of internet path faults on reactive routing.
\newblock In {\em ACM SIGMETRICS Performance Evaluation Review}, volume~31,
  pages 126--137. ACM, 2003.

\bibitem{Freedman05}
Michael Freedman, Mythili Vutukuru, Nick Feamster, and Hari Balakrishnan.
\newblock {Geographic Locality of IP Prefixes}.
\newblock In {\em Internet Measurement Conference (IMC) 2005}, Berkeley, CA,
  October 2005.

\bibitem{Gionis07}
Aristides Gionis, Heikki Mannila, and Panayiotis Tsaparas.
\newblock Clustering aggregation.
\newblock {\em Transactions on Knowledge Discovery from Data}, 1(1), 2007.

\bibitem{Gueye07}
Bamba Gueye, Steve Uhlig, and Serge Fdida.
\newblock Investigating the imprecision of ip block-based geolocation.
\newblock In {\em Proceedings of the 8th International Conference on Passive
  and Active Network Measurement}, PAM'07, pages 237--240, Berlin, Heidelberg,
  2007. Springer-Verlag.

\bibitem{Gursun-thesis}
Gonca G\"{u}rsun.
\newblock Inferring hidden features in the internet.
\newblock In {\em PhD Thesis, Boston University}, 2013.

\bibitem{mytech}
Gonca G\"{u}rsun.
\newblock Routing-aware partitioning of the internet address space for server
  ranking in cdns.
\newblock Technical report, Ozyegin University, 2015.

\bibitem{Gursun12}
Gonca G\"{u}rsun, Natali Ruchansky, Evimaria Terzi, and Mark Crovella.
\newblock Routing state distance: A path-based metric for network analysis.
\newblock In {\em Proceedings of the 2012 ACM Conference on Internet
  Measurement Conference}, IMC '12, pages 239--252, New York, NY, USA, 2012.
  ACM.

\bibitem{Krishnamurthy00}
Balachander Krishnamurthy and Jia Wang.
\newblock On network-aware clustering of web clients.
\newblock In {\em Proceedings of the Conference on Applications, Technologies,
  Architectures, and Protocols for Computer Communication}, SIGCOMM '00, pages
  97--110, New York, NY, USA, 2000. ACM.

\bibitem{Krishnan09}
Rupa Krishnan, Harsha~V. Madhyastha, Sridhar Srinivasan, Sushant Jain, Arvind
  Krishnamurthy, Thomas Anderson, and Jie Gao.
\newblock Moving beyond end-to-end path information to optimize cdn
  performance.
\newblock In {\em Proceedings of the 9th ACM SIGCOMM Conference on Internet
  Measurement Conference}, IMC '09, pages 190--201, New York, NY, USA, 2009.
  ACM.

\bibitem{Labovitz00}
Craig Labovitz, Abha Ahuja, Abhijit Bose, and Farnam Jahanian.
\newblock Delayed internet routing convergence.
\newblock In {\em Proceedings of the Conference on Applications, Technologies,
  Architectures, and Protocols for Computer Communication}, SIGCOMM '00, pages
  175--187, New York, NY, USA, 2000. ACM.

\bibitem{Nygren10}
Erik Nygren, Ramesh~K. Sitaraman, and Jennifer Sun.
\newblock The akamai network: A platform for high-performance internet
  applications.
\newblock {\em SIGOPS Oper. Syst. Rev.}, 44(3):2--19, August 2010.

\bibitem{state-akamai}
The Akamai~State of~the Internet~Report.
\newblock http://www.akamai.com/stateoftheinternet.

\bibitem{Paxson96}
Vern Paxson.
\newblock End-to-end routing behavior in the internet.
\newblock In {\em Conference Proceedings on Applications, Technologies,
  Architectures, and Protocols for Computer Communications}, SIGCOMM '96, pages
  25--38, New York, NY, USA, 1996. ACM.

\bibitem{Poese12}
Ingmar Poese, Benjamin Frank, Bernhard Ager, Georgios Smaragdakis, Steve Uhlig,
  and Anja Feldmann.
\newblock {Improving Content Delivery with PaDIS}.
\newblock {\em IEEE Internet Computing}, 16(3):46--52, May-June 2012.

\bibitem{Poese12-2}
Ingmar Poese, Benjamin Frank, Georgios Smaragdakis, Steve Uhlig, Anja Feldmann,
  and Bruce Maggs.
\newblock Enabling content-aware traffic engineering.
\newblock {\em SIGCOMM Comput. Commun. Rev.}, 42(5):21--28, September 2012.

\bibitem{Poese11}
Ingmar Poese, Steve Uhlig, Mohamed~Ali Kaafar, Benoit Donnet, and Bamba Gueye.
\newblock Ip geolocation databases: Unreliable?
\newblock {\em SIGCOMM Comput. Commun. Rev.}, 41(2):53--56, April 2011.

\bibitem{Rexford02}
Jennifer Rexford, Jia Wang, Zhen Xiao, and Yin Zhang.
\newblock Bgp routing stability of popular destinations.
\newblock In {\em Proceedings of the 2Nd ACM SIGCOMM Workshop on Internet
  Measurment}, IMW '02, pages 197--202, New York, NY, USA, 2002. ACM.

\bibitem{Savage99}
Stefan Savage, Andy Collins, Eric Hoffman, John Snell, and Thomas Anderson.
\newblock The end-to-end effects of internet path selection.
\newblock In {\em Proceedings of the Conference on Applications, Technologies,
  Architectures, and Protocols for Computer Communication}, SIGCOMM '99, pages
  289--299, New York, NY, USA, 1999. ACM.

\bibitem{Streibelt13}
Florian Streibelt, Jan B\"{o}ttger, Nikolaos Chatzis, Georgios Smaragdakis, and
  Anja Feldmann.
\newblock Exploring edns-client-subnet adopters in your free time.
\newblock In {\em Proceedings of the 2013 Conference on Internet Measurement
  Conference}, IMC '13, pages 305--312, New York, NY, USA, 2013. ACM.

\bibitem{vis-akamai}
Visualizing the Internet.
\newblock http://www.akamai.com/html/technology/\\ visualizing$\_$akamai.html.

\bibitem{Wang06}
Feng Wang, Zhuoqing~Morley Mao, Jia Wang, Lixin Gao, and Randy Bush.
\newblock A measurement study on the impact of routing events on end-to-end
  internet path performance.
\newblock {\em SIGCOMM Comput. Commun. Rev.}, 36(4):375--386, August 2006.

\end{thebibliography}

\end{document}